\documentclass[aps,pre,url,twocolumn,superscriptaddress]{revtex4-1}

\usepackage{latexsym}
\usepackage{amssymb,amsmath,amsbsy}
\usepackage{graphicx,color}
\usepackage{bm}
\usepackage{longtable}
\usepackage{epsf}
\usepackage[utf8]{inputenc}
\usepackage{enumerate}
\usepackage[english]{babel}               

\def\eps{\varepsilon}

\def\mv{{\bm v}}

\def\S{\mathcal{S}}
\def\D{\mathcal{D}}
\def\R{{\scriptscriptstyle R}}

\def\eRM{{\mathrm e}}
\def\dRM{{\mathrm d}}

\def\mk{{\bm k}}

\def\mx{{\bm x}}

\def\boldnabla{{\bm \nabla}}
\allowdisplaybreaks

\newcommand{\NB}[1]{\ensuremath{\biggl( #1  \biggl) }}

\newcommand{\fp}[2]{FP$^{\textrm{#1}}_{#2}$}

\begin{document}


\title {Directed percolation process in the presence of velocity
fluctuations: Effect of compressibility and finite correlation time}
\author{N.~V.~Antonov} 
\affiliation{Department of Theoretical Physics, St. Petersburg 
University,\\Ulyanovskaya 1, St. Petersburg, Petrodvorets, 198504 Russia}
\author{M.~Hnati\v{c}}
\affiliation{Institute of Experimental Physics, SAS, 04001 Ko\v{s}ice, Slovakia}
\affiliation{Faculty of Sciences, P.J. \v{S}afarik University, 04154  Ko\v{s}ice, Slovakia}
\affiliation{Bogoliubov Laboratory of Theoretical Physics, JINR, 141980 Dubna, Moscow Region, Russia}
\author{A.~S.~Kapustin}
\affiliation{Department of Theoretical Physics, St. Petersburg 
University,\\Ulyanovskaya 1, St. Petersburg, Petrodvorets, 198504 Russia}
\author{T.~Lu\v{c}ivjansk\'y}
\affiliation{Faculty of Sciences, P.J. \v{S}afarik University, 04154 Ko\v{s}ice, Slovakia}
\affiliation{Fakult\"at f\"ur Physik, Universit\"at Duisburg-Essen, D-47048 Duisburg, Germany} 
\author{L.~Mi\v{z}i\v{s}in}
\affiliation{Faculty of Sciences, P.J. \v{S}afarik University, 04154 Ko\v{s}ice, Slovakia}
\date{\today}
\begin{abstract}
The direct bond percolation process (Gribov process) is studied in the presence of
 random velocity fluctuations generated by the Gaussian self-similar ensemble
 with finite correlation time. We employ the renormalization group in order 
to analyze a combined effect of the compressibility and finite
correlation time on the long-time behavior of the phase transition between an active and
an absorbing
state. The renormalization procedure is performed to the one-loop order. 
Stable fixed points of the renormalization group and their regions of stability are
calculated in the one-loop approximation within the three-parameter 
$(\eps,y,\eta)$-expansion. Different regimes corresponding to the rapid-change limit and frozen velocity field
are discussed, and their fixed points' structure is determined in numerical fashion.
\end{abstract}

\maketitle

\section{Introduction \label{sec:intro}}

%
%
The non-equilibrium physical systems constitute an exciting research topic
to which a lot of effort has been made during last decades
 \cite{Zia95,HHL08,Tauber2014}. 
Absorbing phase transitions between active (fluctuating) and 
inactive (absorbing) states are of particular importance. In these transitions
large scale spatio-temporal
fluctuations of an underlying order parameter take place and the resulting collective
behavior is similar to equilibrium phase transitions.
Such behavior could be observed in many natural phenomena ranging from physics, chemistry,
biology, economy or even sociology. 

A fundamental part of  this type of  systems belongs to the directed percolation (DP)  
 universality class \cite{Stauffer,HHL08}. As pointed by Janssen and Grassberger
 \cite{Janssen81,Grassberger82},  necessary
conditions are: i) a unique absorbing state, ii) short-ranged interactions,
 iii) a positive order parameter and iv) no extra symmetry or additional slow variables.
 Among a few models described within this framework we name population dynamics,
 reaction-diffusion problems \cite{Odor04}, percolation processes \cite{JanTau04}, hadron
 interactions \cite{Cardy80}, etc. 
These models are usually considered without an inclusion of additional
interactions within the mode-mode coupling approach \cite{Hohen77}.
However, in realistic situations impurities and defects,  which are not taken into account
 in the original DP formulation, are expected to cause a change in the
 universal properties of the model. 
 This is believed to be one of the reasons  why there are
not so many direct experimental realizations \cite{RRR03,TKCS07} of the percolation process itself.
 A study of deviations from the ideal situation could proceed in different routes and this 
 still constitutes a topic of an ongoing debate \cite{HHL08}.
A substantial effort has been made  in studying  a long-range interaction using
L{\'e}vy-flight jumps \cite{Jan99,Hin06,Hin07}, effects
 of immunization \cite{Hin01,JanTau04}, mutations \cite{Sarkar15}, feedback of the environment on
 the percolating density \cite{SarkarBasu14}, or in the presence of spatially quenched
 disorder \cite{Janssen97}.
 In general, the novel behavior is observed  with a  possibility that critical
 behavior is lost.
 For example, the presence of a quenched disorder in the latter case causes a shift
 of the critical fixed point to the unphysical region.
 This leads to such interesting phenomena as an activated
 dynamical scaling or Griffiths singularities \cite{MorDic96,CGM98,Vojta05,Vojta06}.

In this paper, we focus on the directed bond percolation process in the presence
of advective velocity fluctuations. Velocity fluctuations are hardly avoidable
in any of experiments. For example,  a vast majority of chemical reactions occurs
at finite temperature, which is inevitably encompassed with the presence of a thermal noise. 
Furthermore, disease spreading and chemical reactions could be affected by the turbulent
advection to a great extent.
Fluid dynamics is in general described by the Navier-Stokes equations \cite{Landau_fluid}.
A general solution of these equations remains an open question
 \cite{Frisch,Monin}. However, to provide more insight we restrict ourselves
 to a more decent problem.
Namely, we assume that the velocity field is given by the Gaussian 
velocity ensemble with prescribed  statistical properties \cite{Kra68,Ant99}.
 Although this assumption appears  as
 oversimplified,  compared to the realistic flows at  the first sight, it nevertheless captures essential
 physical information about advection processes \cite{Kra68,FGV01,turbo}.

Recently, there has been  increased interest in  different advection problems in 
compressible turbulent flows \cite{Benzi09,Pig12,Volk14,depietro15}. These studies show that
compressibility plays a decisive role for population dynamics or chaotic mixing of colloids.
 Our main aim is to investigate an influence of 
 compressibility \cite{AdzAnt98,Ant00}
 on the critical properties of the directed bond percolation process \cite{HHL08}. 
  To this end, 
  the advective field is described by the Kraichnan model with finite correlation time, in which
  not only a solenoidal (incompressible) but also
 a potential (compressible) part of the velocity statistics is involved.
 Note that in our model there is no backward influence of percolating
 field on the velocity fluctuations. In other words, our model corresponds to
 the passive advection of the  reacting scalar field.

A powerful tool for  analysis of the  critical behavior is the renormalization group (RG) 
\cite{Amit,Zinn,Vasiliev} method. 
It constitutes a theoretical framework which allows one to compute universal
quantities in a controllable manner and also to determine universality classes of
the physical system. Here this method is employed in order to
determine the scaling behavior in the vicinity of the phase transition
between the active and absorbing state with an emphasis on a possible type of critical behavior.

The remainder of the paper proceeds as follows. In 
 Sec.~\ref{sec:model}, we introduce a
coarse-grained formulation of the problem, which we reformulate
 into the field-theoretic model. In Sec.~\ref{sec:RG_analysis},  we describe the main steps of the
 perturbative RG procedure. In Sec.~\ref{sec:regimes},  we present 
 analysis of possible regimes involved in the model. We analyze numerically
 and to some extent analytically  fixed points' structure.
 In Sec.~\ref{sec:concl},  we give a concluding summary. 
 Technical details concerning calculation of RG constants
 and functions are presented in Appendix \ref{app:const} and Appendix \ref{app:special}. 
 The coordinates of all fixed points are given in Appendix \ref{app:fixed}.
\section{The model \label{sec:model}}
%
%
A continuum description of DP in terms of a density 
$\psi = \psi(t,\mx)$ of infected individuals typically arises from
a coarse-graining procedure in which a large number of
fast microscopic degrees of freedom are averaged out. A loss of the physical
information is supplemented by a Gaussian noise in a resulting Langevin equation.
Obviously, a correct mathematical description has to be in conformity regarding
the absorbing state condition: $\psi = 0 $ is always a stationary state
and no microscopic fluctuation can change that. 
The coarse grained stochastic equation then reads \cite{JanTau04}
\begin{equation}
  \partial_t {\psi}  = D_0 (\boldnabla^2 - \tau_0)\psi  - 
   \frac{g_0 D_0}{2}\psi^2
  + \xi,
  \label{eq:basic}
\end{equation}
where $\xi$ denotes the noise term, $\partial_t = \partial / \partial t$ is
the time derivative, $\boldnabla^2$ is  the Laplace operator, $D_0$ 
is the diffusion constant, $g_0$ is the coupling constant and $\tau_0$ measures
 a deviation from the threshold value for injected probability. It can be thought
 as an analog to the temperature  variable in the standard $\varphi^4-$theory 
 \cite{JanTau04,Zinn}.
 Due to  dimensional reasons,  we have extracted the 
 dimensional part from the interaction term (See later Sec. \ref{sec:can_dim}). 
Here and henceforth 
we distinguish between
unrenormalized (with the subscript ``0'') quantities and renormalized terms
(without the subscript ``0'').
 The renormalized fields will be later denoted by the subscript $R$.

 It can be rigorously proven \cite{Janssen81} that
the Langevin equation (\ref{eq:basic}) captures the gross properties
 of the percolation process and contains essential physical information about the
 large-scale behavior of the non-equilibrium phase 
 transition between the active $(\psi > 0)$ and the absorbing state $(\psi = 0)$.    
The Gaussian noise term $\xi$ with zero mean
 has to satisfy the absorbing state condition. Its
 correlation function can be chosen in the following form
\begin{equation}
   \langle \xi(t_1,\mx_1) \xi(t_2,\mx_2) \rangle = g_0 D_0 \psi(t_1,\mx_1) 
   \delta(t_1-t_2) \delta^{(d)}(\mx_1 - \mx_2)
   \label{eq:noise_correl}
\end{equation}
up to irrelevant contributions \cite{Tauber2014}. Here $\delta^{(d)}(\mx) $ 
is the $d$-dimensional generalization of the standard Dirac $\delta(x)$-function.
%
%

A further step consists in  incorporating  of the velocity fluctuations into
the model (\ref{eq:basic}). The 
 standard route \cite{Landau_fluid} based on 
   the replacement $\partial_t$ by the Lagrangian derivative $\partial_t +({\bm v}\cdot{\bm \nabla})$ 
   is not sufficient due to the assumed compressibility. As shown in \cite{AntKap10},  the following
   replacement is then adequate
\begin{equation}
  \partial_t \rightarrow \partial_t +({\bm v}\cdot\boldnabla)+a_0 ({\bm \nabla}\cdot{\bm v}),
  \label{eq:subs}
\end{equation}
where $a_0$ is an additional positive parameter, whose significance will be discussed later.
%
%
Note that the last term in (\ref{eq:subs}) contains
a divergence of the velocity field $\bm v$ and thus ${\bm \nabla}$ operator does not act on
what could possibly follow.
%
%

Following  \cite{Ant00},  we
 consider the velocity field to be a random Gaussian variable with zero mean and 
 a translationally invariant correlator given as follows:
\begin{equation}
  \langle v_i(t,{\bm x}) v_j (0,{\bm 0}) \rangle =
  \int \frac{{\mathrm d} \omega}{2\pi}
  \int \frac{{\mathrm d}^d {\bm k}}{(2\pi)^d} 
  D_v (\omega,\mk) {\mathrm e}^{-i\omega  t  +{\bm k}\cdot {\bm x}},
  \label{eq:vel_correl}
\end{equation}
where the kernel function $D_v(\omega,\mk)$ takes the form
%
\begin{equation}
  D_v (\omega,\mk) = [P_{ij}^{k} + \alpha Q_{ij}^{k}]
  \frac{g_{10} u_{10} D_0^3 k^{4-d-y-\eta}}{\omega^2 + u_{10}^2 D_0^2 (k^{2-\eta})^2}.
  \label{eq:kernelD}
\end{equation}
Here, $P_{ij}^k = \delta_{ij}-k_ik_j/k^2$ is a transverse  and $Q_{ij}^k=k_ik_j/k^2$ a longitudinal
projection
operator, $k=|\mk|$, and $d$ is the 
dimensionality of the $\mx$ space.
A positive parameter $\alpha>0$ can be interpreted as the simplest possible
deviation \cite{AdzAnt98} from the incompressibility condition ${\bm \nabla}\cdot {\bm v} = 0$.
The incompressible case, $\alpha=0$, was analyzed in previous 
works \cite{AntKap08,AntKap10,Ant11,SarkarBasu12,DP13}.
The coupling constant $g_{10}$ and the exponent $y$ describe the equal-time velocity
correlator or, equivalently, the energy spectrum \cite{Ant99,Ant00,Frisch} of the velocity
fluctuations. The constant $u_{10}>0$ and the exponent $\eta$ are related
to the characteristic frequency $\omega \simeq u_{10} D_0 k^{2-\eta}$ of the mode with
the wavelength $k$.

The momentum integral in (\ref{eq:vel_correl}) has an infrared (IR) cutoff
at $k = m$, where $m \sim 1/L $ is the
reciprocal of the integral scale $L$. A precise form of the cutoff \cite{turbo,Ant06} is actually
unimportant and its role is to provide us with IR regularization.
Further, dimensional considerations show that the bare coupling constants $g_{10}$ and $u_{10}$
are related to the characteristic UV momentum scale $\Lambda$ by
\begin{equation}
   g_{10} \simeq \Lambda^y,\quad
   u_{10} \simeq \Lambda^\eta.
\end{equation}
The choice $y = 8/3$ gives the famous Kolmogorov ``five-thirds'' law for
the spatial velocity correlations, and $\eta=4/3$ corresponds to the Kolmogorov
frequency \cite{Frisch}.

The exponents $y$ and $\eta$ are analogous to the standard expansion parameter
$\eps = 4-d$ in the static critical phenomena.
It can be shown that the upper critical dimension of the pure
percolation problem \cite{JanTau04} is also $d_c = 4$.
Therefore, we retain the standard notation for the exponent $\eps$.
According to the general rules \cite{Vasiliev} of the  RG approach,  we formally assume
that  the exponents $\eps,y$ and $\eta$ are of the same order of magnitude and
 constitute small expansion parameters of perturbation theory.

%
%
%
The kernel function in (\ref{eq:kernelD}) is chosen in a quite
general form
 and as such it contains various special limits. 
 They simplify numerical analysis of the resulting equations and
 allows us to gain a deeper physical insight into the model.
 Possible limiting cases are
\begin{enumerate}[i)]
  \item The rapid-change model, which corresponds
	to the limit $u_{10} \rightarrow \infty, g_{10}' \equiv g_{10}/u_{10} = const$.
	Then for the kernel function we have
	\begin{equation}
	   D_v(\omega,{\mk}) \propto g_{10}' D_0 k^{-d-y+\eta}
	   \label{eq:rch_limit}
	\end{equation}
	 and obviously the velocity correlator is $\delta-$correlated in a time
	 variable.	
  \item The frozen velocity field, which arises in the limit $u_{10} \rightarrow 0$ 
	and the kernel function corresponds to 
	\begin{equation}
	   D_{v}(\omega,{\mk}) \propto g_0 D_0^2 \pi \delta(\omega) k^{2-d-y}.
	   \label{eq:fvf_limit}
	\end{equation}
  \item The purely potential velocity field , which is obtained for 
	$\alpha\rightarrow\infty$ with $\alpha g_{10}=$constant.
	This limit is similar to
	the model of random walks in a random environment with long-range
	correlations \cite{HonKar88,Bou90}.
  \item The turbulent advection, for which the $y=2\eta=8/3$. This choice mimics
	properties of the genuine turbulence and leads to the celebrated Kolmogorov
	scaling \cite{Frisch}.
\end{enumerate}

%
%
For an effective use of the RG method it is advantageous
to rewrite the stochastic problem (\ref{eq:basic}-\ref{eq:kernelD})
into the field-theoretic formulation. This could be achieved
 in the standard fashion \cite{Janssen76,deDom76,Janssen79}
and the resulting dynamic functional is 
\begin{equation}
   \S[\varphi] = \S_{ \text{diff}}[\varphi]
   + \S_{\text{vel}}[\varphi]
   + \S_{\text{int}}[\varphi], 
   \label{eq:bare_act}
\end{equation}
where $\varphi=\{\tilde{\psi},\psi,\mv \}$ stands for the complete set of fields
and $\tilde{\psi}$ is the auxiliary (Martin-Siggia-Rose) response field \cite{MSR73}. 
The first term represents a free part of the equation (\ref{eq:basic})
and is given by the following expression:
\begin{equation}
  \S_{ \text{diff}}[\varphi] =  
  \int \dRM t \int \dRM^{d} \mx \biggl\{
  \tilde{\psi}[
  \partial_t - D_0\boldnabla^2+D_0\tau_0
  ]\psi \biggl\}.
  \label{eq:act_diffuse}
\end{equation}
Since the velocity fluctuations are governed by the Gaussian statistics, the
 corresponding averaging procedure is performed with the quadratic functional 
\begin{align}
  \S_{\text{vel}}[\mv] & = \frac{1}{2} 
  \int \dRM t_1 \int \dRM t_2 
  \int \dRM^d \mx_1 \int \dRM^d \mx_2
  \mbox{ }
  \mv_i(t_1,x_1) \nonumber \\
  &  D_{ij}^{-1}(t_1-t_2,\mx_1-\mx_2) \mv_j(t_2,\mx_2),
  \label{eq:vel_action}
\end{align}
where $D_{ij}^{-1}$ is the kernel of the inverse linear operation in (\ref{eq:vel_correl}).
The final interaction part can be written as
\begin{align}
  \S_{\text{int}}[\varphi] & = 
  \int \dRM t \int \dRM^{d} \mx
  \biggl\{  
  \frac{D_0\lambda_0}{2} [\psi-\tilde{\psi}
  ]\tilde{\psi}\psi
  -\frac{u_{20}}{2D_0} \tilde{\psi} \psi
  \mv^2 \nonumber \\
  & +  \tilde{\psi} (\mv\cdot\boldnabla) \psi 
  +a_0 \tilde{\psi} (\boldnabla\cdot\mv)\psi
  \biggl\}.
  \label{eq:inter_act}
\end{align}

All but the third term in (\ref{eq:inter_act}) directly stem from the nonlinear
terms in (\ref{eq:basic}) and (\ref{eq:subs}).
The third term proportional to $\propto \tilde{\psi}\psi\mv^2$ deserves a special consideration. 
The presence of this term is prohibited in the original Kraichnan model due
to the underlying Galilean invariance. However, in our case the general form of
the velocity kernel function does not lead to such restriction. Moreover, by direct
inspection of the perturbative expansion, one can show that this kind of  term is indeed generated
under RG transformation (consider second Feynman graph in the expression (\ref{eq:exp_ppvv})).
This term was considered
for the first time in our previous work \cite{DP13}, where the incompressible case
is analyzed.

Let us also note that for the linear advection-diffusion equation \cite{Ant00,Landau_fluid}, the
choice $a_0=1$ corresponds to the conserved quantity $\psi$ (advection of a
density field), whereas for the choice $a_0=0$ the conserved
quantity is $\tilde{\psi}$ (advection of a tracer field).
 From the point of view of the renormalization 
group, the introduction of $a_0$ is necessary,
because it ensures multiplicative renormalizability of the model \cite{AntKap10}.

In principle,  basic ingredients of any stochastic theory, correlation and 
response functions of the concentration 
field $\psi(t,\mx)$, can  be
computed as functional averages with respect to the weight functional $\exp(-\S)$
with action (\ref{eq:bare_act}).
Further, the field-theoretic formulation summarized in (\ref{eq:act_diffuse})-(\ref{eq:inter_act})
 has an additional advantage to be amenable to the full machinery of (quantum) field theory
 \cite{Zinn,Vasiliev}.
In the subsequent section,  we apply the RG perturbative technique \cite{Vasiliev} that allows us
to study the model in the vicinity of its upper critical dimension $d_c=4$.
\section{Renormalization group analysis \label{sec:RG_analysis}}
%
%
%

An important goal of statistical theories is the determination of correlation and response functions
(usually called Green functions) of the dynamical fields as functions of the space-time coordinates. 
Traditionally, these functions are represented in the  form of sums over the Feynman diagrams 
\cite{Vasiliev,Zinn}.
The functional formulation provides a convenient theoretical framework
suitable for applying methods of quantum field theory.
Using the RG method \cite{WilKog74,Zinn}  it is possible to determine the infrared (IR)
asymptotic (large spatial and time scales) behavior of the correlation 
functions. A proper renormalization procedure is needed for the elimination of 
ultraviolet (UV) divergences.
There are various renormalization prescriptions applicable for such 
 problem, each with its own
advantages \cite{Zinn}. In this work,  we employ the
 minimal subtraction (${\text{MS}}$) scheme. 
UV divergences manifest themselves
in the form of poles in the small expansion parameters,  and the minimal subtraction scheme
is characterized by discarding all finite parts of the Feynman graphs in the
calculation of the renormalization constants.
In the vicinity of critical points large fluctuations on all spatio-temporal scales dominate
the behavior of the system, which in turn results in the divergences in the Feynman graphs.
The resulting RG functions satisfy certain differential equations and their analysis
 provides us with an efficient computational technique for estimation of universal quantities.
\subsection{Canonical dimensions \label{sec:can_dim}}
In order to apply the dimensional regularization for  evaluation 
of renormalization constants, an analysis of possible superficial
divergences has to be performed.
For translationally invariant
systems, it  is sufficient to analyze only 1-particle irreducible (1PI)
graphs \cite{Zinn,Amit}.
In contrast to static models, dynamic models  \cite{Vasiliev,Tauber2014} 
contain two independent scales: a frequency scale 
 $d^\omega_Q $ and a momentum scale $d^k_Q$ for each quantity $Q$.
The corresponding dimensions are found using the 
standard normalization conditions 
\begin{align}
  &  d_k^k = - d^k_x =1,\quad
  & d^k_\omega & = d_t^k = 0,\nonumber\\
  & d_k^\omega = d^\omega_x = 0,\quad
  & d^\omega_\omega & = -d_t^\omega = 1
  \label{eq:def_normal}
\end{align}
together with a condition for  field-theoretic action to be a dimensionless quantity.
Using the quantities  $d^\omega_Q$ and $d_Q^k$,
the total canonical dimension $d_Q$,
\begin{equation}
   d_Q = d_Q^k + 2d_Q^\omega
\end{equation}
can be introduced,
whose precise form is obtained from a comparison of the IR most
relevant terms ($\partial_t \propto \boldnabla^2$) in the action (\ref{eq:act_diffuse}).
The total dimension $d_Q$ for the dynamical models
plays the same role as the conventional (momentum) dimension does in static problems.
The dimensions of all quantities for the model are summarized in Table \ref{tab:canon}.
It follows that the model is logarithmic (when coupling constants
are dimensionless) at $\eps = y = \eta = 0$, and the UV divergences are
in principle realized as poles in these parameters.
\begin{table}
\begin{tabular}{| c | c | c | c | c| c | c | c | c | }
  \hline\noalign{\smallskip}
  $Q$ & $\psi,\tilde{\psi}$ & ${\mv}$ & $D_0$ & $\tau_0$ & $g_{10}$ & $\lambda_0 $  
  & $u_{10}$  & $u_{20},a_0,\alpha$
 \\  \noalign{\smallskip}\hline\noalign{\smallskip}
  $d_Q^k$ & $d/2$ & $-1$ & $-2$ & $2$ & $y$ & $\eps/2$
  & $\eta$  & $0$ 
  \\  \noalign{\smallskip}\hline\noalign{\smallskip}
  $d^\omega_Q$ & 0 & $1$ & $1$ & $0$ & $0$ & $0$
  & $0$  & $0$ 
  \\  \noalign{\smallskip}\hline\noalign{\smallskip}
  $d_Q$ & $d/2$ & $1$ & $0$ & $2$ & $y$ & $\eps/2$
  & $\eta$  & $0$ 
  \\ \noalign{\smallskip}\hline    
\end{tabular}
  \caption{Canonical dimensions of the bare fields and bare parameters 
	  for the model (\ref{eq:act_diffuse})-(\ref{eq:inter_act}).  }
  \label{tab:canon}
\end{table}
The total canonical dimension of an arbitrary $1-${ irreducible} Green function
is given by the relation
\begin{equation}
   d_\Gamma = d^k_\Gamma + 2 d^\omega_\Gamma = d + 2 - \sum_\varphi N_\varphi d_\varphi, \quad 
   \varphi\in\{\tilde{\psi}, \psi, \mv \}.
   \label{eq:def_dim}
\end{equation}
The total dimension $d_\Gamma$ in the logarithmic theory is the formal degree of the 
UV divergence $\delta_\Gamma = d_\Gamma |_{\eps=y=\eta=0}$. 
Superficial UV divergences, whose removal requires counterterms, could be
present only in those functions $\Gamma$ for which $\delta_\Gamma$ is
a non-negative integer \cite{Vasiliev}.

\begin{table}
  \begin{tabular}{|c|c|c|c|c|c|c|c|c|}
    \hline\noalign{\smallskip}
    $\Gamma_{1-ir}$ &  $\Gamma_{\tilde{\psi} \psi} $& $\Gamma_{\tilde{\psi} \psi \mv }$
                  & $\Gamma_{\tilde{\psi}^2 \psi} $ & $ \Gamma_{\tilde{\psi}\psi^2} $
                  & $\Gamma_{\tilde{\psi}\psi \mv^2}$                                    
   \\
    \noalign{\smallskip}\hline\noalign{\smallskip}
    $d_\Gamma$ & $2$ & $1$ & $\eps/2$ & $\eps/2$ & $0$
               \\
     \noalign{\smallskip}\hline\noalign{\smallskip}
     $\delta_\Gamma$ & $2$ & $1$ & $0$ & $0$ & $0$
               \\ \hline
  \end{tabular}
    \caption{Canonical dimensions for the (1PI) divergent Green functions of the model.}
  \label{tab:canon_green}
\end{table}
Dimensional analysis should be augmented by certain additional considerations.
In dynamical models of the type (\ref{eq:act_diffuse})-(\ref{eq:inter_act}), all
the 1-irreducible diagrams without the
fields $\tilde{\psi}$ vanish, and it is sufficient to consider the functions with 
$N_{\tilde{\psi}} \ge 1$. As was shown in \cite{AntKap10},  the 
rapidity symmetry (\ref{eq:time_sym})
requires also $N_{\psi} \ge 1$ to hold. Using these considerations
together with  relation (\ref{eq:def_dim}), possible UV divergent structures
are expected only for the 1PI Green functions listed in
Table \ref{tab:canon_green}.
\subsection{Computation of the RG constants \label{sec:RG_const}}
%
%
In this section, the  main steps of the perturbative RG approach are
summarized, deferring the explicit results of
the RG constants and RG functions (anomalous dimensions and
beta functions) to Appendices \ref{app:const} and \ref{app:special}.

A starting point of the perturbation theory is a free part of the action
given by  expressions (\ref{eq:act_diffuse}) and (\ref{eq:vel_action}).
By graphical means, they are represented as lines in the Feynman diagrams, 
whereas the non-linear terms  in (\ref{eq:inter_act}) correspond to vertices
connected by the  lines.

For the calculation of the  RG constants  we have employed dimensional regularization in the
combination with the MS scheme \cite{Zinn}. 
Since the finite correlated case involves two different dispersion laws: 
$\omega\propto k^2$ for the scalar and
$\omega\propto k^{2-\eta}$
for the velocity fields, the calculations for the renormalization constants become
rather cumbersome already in the one-loop approximation \cite{Ant99,Ant00}.
However, it was shown \cite{AdzAntHon02} that to the two-loop order it is sufficient to
consider the choice $\eta = 0$. This significantly simplifies practical calculations
and as can be seen in (\ref{eq:RG_inverse1}), the only poles to the one-loop order are of
 two types: either $1/\eps$ or $1/y$.
This simple picture pertains only to the lowest orders in a
perturbation scheme. In higher order terms, poles in the form
of general linear combinations in $\eps,\eta$ and $y$ are expected to arise.

\begin{figure}
   \includegraphics[width=3cm]{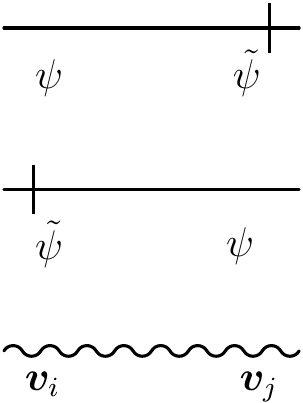}
   \caption{Diagrammatic representation of the bare
   propagators. The time flows from right to left.}
   \label{fig:prop}
\end{figure}

The perturbation theory of the model (\ref{eq:bare_act}) is amenable
to the standard Feynman diagrammatic expansion \cite{Amit,Zinn,Vasiliev}.
The inverse matrix of the quadratic part in the actions determines a form of the bare propagators.
 The propagators are presented in the wave-number-frequency representation, which is
for the translationally invariant systems the most convenient way for doing explicit calculations.
The bare propagators are easily read off from the Gaussian part of the model
given by (\ref{eq:act_diffuse}) and (\ref{eq:vel_action}), respectively. 
Their graphical representation
is depicted in Fig.~\ref{fig:prop}. The corresponding algebraic expressions 
can be easily read off and in the frequency-momentum representation are
given by
\begin{align}
   \langle {\psi} \tilde{\psi}\rangle_0 
   & = 
    \langle \tilde{\psi} \psi\rangle_0^* 
    =
   \frac{1}{-i\omega + D_0(k^2+\tau_0)}, \label{eq:prop1a} \\
   \langle \mv \mv \rangle_0 
   & =  
   [P_{ij}^{k} + \alpha Q_{ij}^{k}]
   \frac{g_{10} u_{10} D_0^3 k^{4-d-y-\eta}}{\omega^2 + u_{10}^2 D_0^2 (k^{2-\eta})^2}
   \label{eq:prop2a}
\end{align}
or in the time-momentum representation as
 \begin{align}  
    \langle {\psi} \tilde{\psi}\rangle_0 
   & = \theta(t) \exp(-D_0[k^2+\tau_0]t), 
   \\
    \langle \tilde{\psi} \psi\rangle_0 
    & = \theta(-t) 
    \exp(D_0[k^2+\tau_0]t), \label{eq:prop1b} \\
   \langle \mv \mv \rangle_0  
   & =  
   [P_{ij}^{k} + \alpha Q_{ij}^{k}]
   \frac{g_{10}D_0^2 }{k^{d+y-2}}\eRM^{-u_{10}D_0 k^{2-\eta}|t|},
   \label{eq:prop2b}
\end{align}
where $\theta(t)$ is the Heaviside step function.

\begin{figure}
   \includegraphics[width=7cm]{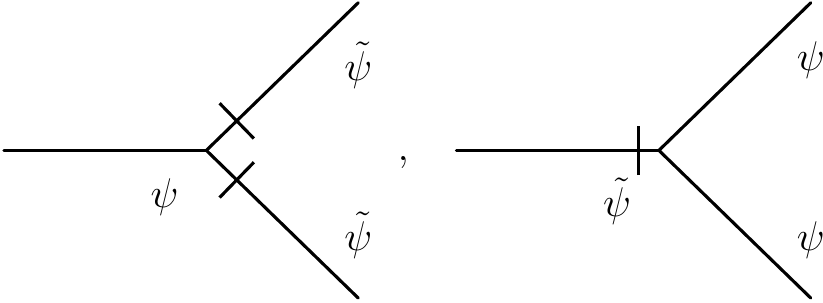}
   \caption{Diagrammatic representation of the interaction vertices 
   describing an  ideal directed bond percolation process.}
   \label{fig:vert1}
\end{figure}

\begin{figure}
   \includegraphics[width=7cm]{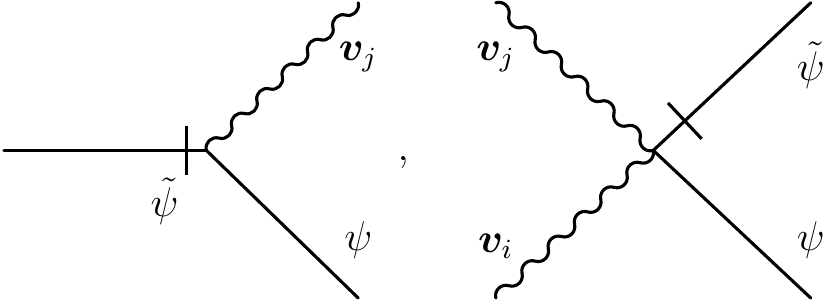}
   \caption{Interaction vertices describing the influence of the advecting velocity field
   with the  order parameter fluctuations.
   }
   \label{fig:vert2}
\end{figure}

The interaction vertices from the nonlinear part
(\ref{eq:inter_act}) describe
 the fluctuation effects connected with the 
 percolation process  itself,
 advection of the  concentration field and the interactions between
 the velocity components.
With every such vertex
the following algebraic factor 
\begin{equation*}
  V_N(x_1,\ldots,x_N;\varphi) = 
  \frac{\delta^N \S_{\text{int}}[\varphi]}{\delta\varphi(x_1)\ldots\delta\varphi(x_N)},
  \quad
  \varphi \in\{\tilde{\psi},\psi,\mv \}
  \label{eq:ver_factor}
\end{equation*}
is associated \cite{Vasiliev}. 
In our model there are four different interaction vertices, which are
graphically depicted in Fig.~\ref{fig:vert1} and Fig.~\ref{fig:vert2}, respectively.
The corresponding vertex factors are
\begin{align}
   & V_{\tilde{\psi}{\psi} \psi} =
       - V_{\tilde{\psi}\tilde{\psi} \psi} = D_0\lambda_0,\\
   & V_{\tilde{\psi}\psi\mv} =  ik_j +i a_0 q_j,\\
   & V_{\tilde{\psi}\psi\mv\mv} = - \frac{u_{20}}{D_0}\delta_{ij}.
   \label{eq:factors}
\end{align}
In the expression for $V_{\tilde{\psi}\psi\mv}$, we have adopted the following convention:
$k_j$ is the momentum of the field $\psi$ and $q_j$
is the momentum of the velocity field $\mv$.
The presence of the interaction vertex $V_{\tilde{\psi}\psi\mv\mv}$ leads
to the proliferation of the new Feynman graphs (see Appendix \ref{app:const}), which
were absent in the previous studies \cite{AntKap08,AntKap10,DP13}. 
%

By direct inspection of the Feynman diagrams one can 
observe that the real expansion parameter is 
rather $\lambda_0^2$ than $\lambda_0$. This is a direct consequence of the duality symmetry 
\cite{JanTau04}
of the action for the pure percolation problem with respect to time inversion
\begin{equation}
  \psi(t,\mx) \rightarrow -\tilde{\psi}(-t,\mx),\quad
  \tilde{\psi}(t,\mx) \rightarrow -\psi(-t,\mx).
  \label{eq:time_sym}
\end{equation}
Therefore, we introduce a
new charge $g_{20}$ via the relation
\begin{equation}
  g_{20} = \lambda_0^2
  \label{eq:new_g2}
\end{equation}
and express the perturbation calculation in terms of this parameter.

In the presence of compressible velocity field the transformation
(\ref{eq:time_sym})
has to be augmented by the transformation 
\begin{equation}
  a_0 \rightarrow 1-a_0,
  \label{eq:supple}
\end{equation}
as can be easily seen by inserting (\ref{eq:time_sym}) in (\ref{eq:inter_act})
and performing integration by parts. 

With the help of Table \ref{tab:canon} the renormalized parameters can be introduced
in the following manner:
\begin{align}
   \label{eq:RGconst}
   &D_0 = D Z_D, &\tau_0& = \tau Z_\tau + \tau_c,
     &a_0& = a Z_a,  
     \nonumber \\ 
   &g_{10} = g_{1} \mu^{y} Z_{g_1}, &u_{10}& = u_1 \mu^\eta Z_{u_1},
   &\lambda_0& = \lambda \mu^{\eps/2} Z_\lambda, \nonumber
   \\
   &g_{20}=g_2 \mu^{\eps} Z_{g_2}, &u_{20}& = u_2 Z_{u_2}, 
\end{align}
where $\mu$ is the reference mass scale in the MS scheme~\cite{Zinn}. Note that
the term $\tau_c$ is a non-perturbative effect \cite{Sym73,Schloms89}, which
is not captured by the  dimensional regularization.
The renormalization prescription (\ref{eq:RGconst}) together with the 
renormalization of fields 
\begin{equation}
 \tilde{\psi} = Z_{\tilde{\psi}} \tilde\psi_{\R},\quad
 \psi = Z_\psi \psi_{ \R},\quad
 \mv = Z_v{\mv}_{\R} 
  \label{eq:RGfields}
\end{equation}
is sufficient for obtaining a fully renormalized theory.
Thus,  the total renormalized action for
the renormalized fields $\varphi_R \equiv \{\tilde\psi_{\R}, \psi_{\R}, \mv_{\R} \}$ 
  can be written in a compact form
\begin{align}
  \S_R[\varphi_\R] & = \int \dRM t \int \dRM^d \mx
  \biggl\{ \tilde{\psi}_\R \biggl[
  Z_1\partial_t - Z_2 D\nabla^2 + Z_3 D\tau \nonumber \\
  & + Z_4 (\bm{v}_\R\cdot \bm{\nabla}) 
    +  a Z_5 (\bm{\nabla}\cdot\bm{v}_\R) 
   \biggl] \psi_\R
  -  \frac{D\lambda}{2}[Z_6 \tilde{\psi}_\R \nonumber \\
  & - Z_7
  \psi_\R]\tilde{\psi_\R}\psi_\R - Z_8\frac{u_2}{2D}\tilde{\psi_\R} \psi_\R \bm{v}^2_\R \biggl\}
  + \nonumber\\
  &
  \frac{1}{2} 
  \int \dRM t_1 \int \dRM t_2 
  \int \dRM^d \mx_1 \int \dRM^d \mx_2
  \mbox{ }
  \mv_{Ri}(t_1,x_1) \nonumber \\
  &  D_{Rij}^{-1}(t_1-t_2,\mx_1-\mx_2) \mv_{Rj}(t_2,\mx_2).
  \nonumber\\
   \label{eq:renorm_action}
\end{align}
The latter term is a renormalized version of (\ref{eq:vel_action}).
The relations between the
renormalization constants follow directly from the action (\ref{eq:renorm_action}) 
\begin{align}
  \label{eq:RG_direct}
  &Z_1  = Z_\psi Z_{\tilde{\psi}}, 
  & Z_2 & =  Z_\psi Z_{\tilde{\psi}} Z_D, \nonumber \\
  &Z_3  = Z_\psi Z_{\tilde{\psi}} Z_D Z_\tau,
  & Z_4 & =  Z_\psi Z_{\tilde{\psi}} Z_v, \nonumber \\
  &Z_5  = Z_\psi Z_{\tilde{\psi}} Z_v Z_a, 
  & Z_6 & =  Z_\psi Z_{\tilde{\psi}}^2 Z_D Z_\lambda, \nonumber   \\
  &Z_7  = Z_\psi^2 Z_{\tilde{\psi}} Z_D Z_\lambda, 
  & Z_8 & =  Z_\psi Z_{\tilde{\psi}} Z_v^2 Z_{u_2}Z_D^{-1}. 
\end{align}
The theory is made UV finite through the appropriate choice
of the RG constants $Z_1,\ldots,Z_8$. Afterwards,   relations (\ref{eq:RG_direct}) yield
 the corresponding RG constants for the fields and parameters
appearing in relations (\ref{eq:RGconst}). 
The explicit results for the RG constants are given in Appendix \ref{app:const}.

According to the general rules of the RG method \cite{Vasiliev}, the nonlocal term in action
(\ref{eq:renorm_action})
should not be renormalized. From the inspection of the kernel function (\ref{eq:kernelD})
 two additional relations
\begin{equation}
  1  = Z_{u_1} Z_D, \quad
  1  =  Z_{u_1} Z_{g_1}Z_D^3 Z_v^{-2}
  \label{eq:RG_special}  
\end{equation}
 follow, which have to be satisfied to all orders in the perturbation scheme.

\section{Fixed points and scaling regimes \label{sec:regimes}}
%
%
%

Once the renormalization procedure to a given order of perturbation
scheme is performed, we can find the scaling 
behavior in the infrared IR limit by studying the flow as $\mu\rightarrow 0$.
According to the general statement of the RG theory \cite{Amit,Vasiliev}, a
possible IR asymptotic behavior is 
governed by the fixed point (FP) of the beta-functions. 
All fixed points 
can be found from a requirement that all beta-functions of the model simultaneously vanish
\begin{align}
  &\beta_{g_1} (g^{*}) =\beta_{g_2} (g^{*})= \beta_{u_1}
  (g^{*})=\beta_{u_2} (g^{*})=\beta_{a} (g^{*})=0,\nonumber\\
  \label{eq:gen_beta}
\end{align}
where $g^{*}$ stands for an entire set of charges $\{g_1^{*},g_2^{*},u_1^{*},u_2^{*},a^{*}\}$.
In what follows,  the asterisk will always refer to coordinates of
some fixed point.
Whether the given FP could be realized in physical systems (IR stable) or not 
(IR unstable)
is determined by eigenvalues of the matrix $\Omega=\{\Omega_{ij}\}$ with
the elements
\begin{equation}
   \Omega_{ij} = \frac{\partial \beta_i}{\partial g_j},
   \label{eq:matrix}
\end{equation}
where $\beta_i$ is a full set of beta-functions and $g_j$ is 
the full set of charges $\{ g_1 ,g_2 ,u_1 ,u_2 ,a \}$.
For the IR stable FP the real parts of the eigenvalues of the matrix $\Omega$ have to be
strictly positive. In general, these conditions determine a region of stability for
the given FP in terms of $\eps,\eta$ and $y$.    

Furthermore, to obtain the RG equation,  one can exploit a fact that the bare Green functions
are independent 
of the  momentum scale $\mu$ \cite{Amit}.
Applying the differential operator $\mu\partial_\mu$ at the fixed bare quantities leads to
 the following equation for the renormalized Green function $G_R$ 
\begin{equation}
  \{ \D_{\text{RG}} + N_{\psi} \gamma_{\psi} +
   N_{\tilde{\psi}} \gamma_{\tilde{\psi}} +N_\mv \gamma_{\mv}  \} 
  G_R(e, \mu, \dots)=0,
  \label{eq:basic_RG}
\end{equation}
where $G_R$ is a function of the full set  $e$ of renormalized counterparts to the bare
 parameters $e_0 =\{D_0, \tau_0, u_{10},u_{20}, g_{10}, g_{20}, a_0 \}$,
 the reference mass scale $\mu$ and other parameters, e.g. spatial and time variables. 
 The RG operator $\D_{\text{RG}}$ is given by
\begin{equation}
   \D_{\text{RG}}\equiv \mu\partial_\mu|_0  = \mu \partial_{\mu} + 
            \sum_{g} \beta_g \partial_g 
	    - \gamma_D \D_D - \gamma_{\tau} \D_{\tau},
   \label{eq:RG_equation}	    
\end{equation}
where $g\in\{g_1,g_2,u_1,u_2,a \}$, $\D_x = x \partial_x$ for any variable $x$, $\dots|_0$
stands for fixed bare parameters
and $\gamma_x$ are the so-called anomalous dimensions of the quantity $x$  defined as 
\begin{equation}
  \label{eq:def_gamma}
  \gamma_x \equiv \mu\partial_\mu \ln  Z_x |_0.
\end{equation}
The beta-functions, which express the flows of parameters under the RG 
transformation \cite{Amit}, are defined through
\begin{equation}
  \label{eq:def_beta}
  \beta_g = \mu \partial_{\mu} g |_{0}.  
\end{equation}
Applying 
this definition to  relations (\ref{eq:RGconst}) yields
\begin{align}
   \beta_{g_1} &= g_1 (-y + 2\gamma_D-2\gamma_v), \qquad
      \beta_{g_2} = g_2 (-\eps -\gamma_{g_2}), \nonumber \\
   \beta_{u_1} &= u_1(-\eta +\gamma_D), \hskip1.8cm
      \beta_{u_2} = - u_2 \gamma_{u_2}, \nonumber \\
   \beta_{a} &= - a \gamma_{a}.
  \label{eq:beta_functions}
\end{align}
The last equation suggests that for the fixed points' equation
$\beta_a(g^*) = 0$ either $a=0$ or $a\neq 0$ has to be satisfied. However, as the
explicit results (\ref{eq:gen_anom_charges}) show,  this is not true (parameter
$a$ appears also in the denominator of $\gamma_a$) and 
the right-hand side of $\beta_a$ has to be considered as a whole expression.
 A similar reasoning also applies for the function $\beta_{u_2}$.
 
It turns out that for some fixed points the computation of the eigenvalues
of the matrix (\ref{eq:matrix}) is cumbersome and rather unpractical. In those
cases it is possible to obtain information about the stability from
analyzing  RG flow equations \cite{Vasiliev}.
 Its essential idea is to study a set of invariant charges $\overline{g} = \overline{g}(s,g)$
 with the initial data $\overline{g} |_{s=1} = g$. The parameter $s$
 stands for a scaling parameter and one is interested in the behavior of charges in 
 the limit $s\rightarrow 0$. The evolution of invariant charges is given by
 the equation
 \begin{equation}
   \mathcal{D}_s \overline{g} = \beta(\overline{g}).
   \label{eq:invariant_chrg}
 \end{equation}

The very existence of IR stable solutions of the RG equations leads
to the existence of the scaling behavior of Green functions. In dynamical
models, critical dimensions of the quantity $Q$ is given by the  relations
\begin{equation}
   \Delta_Q = d^k_Q + \Delta_\omega d^\omega_Q + \gamma^*_Q, \quad
   \Delta_\omega = 2 - \gamma_D^*.
   \label{eq:critic_dims}
\end{equation}
The $d_k^Q$ and $d_\omega^Q$ are canonical dimensions of the quantity
$Q$ calculated with the help of Tab. \ref{tab:canon_green}, $\gamma_Q^*$ is the value
of its anomalous dimension. Using Eqs. (\ref{eq:critic_dims}) we obtain the following relations
\begin{align}
   & \Delta_{\tilde{\psi}} = \frac{d}{2} + \gamma_{\tilde{\psi}},\quad
   \Delta_{\psi} = \frac{d}{2} + \gamma_\psi,\quad
   \Delta_\tau = 2 + \gamma_\tau^*.
\end{align}

Important information about the  physical system can be read out from the behavior
of correlation functions, which can be expressed
in terms of the  cumulant Green functions. In the percolation problems one is typically
interested \cite{HHL08,JanTau04} in the behavior of the following functions
\begin{enumerate}[a)]
  \item The number $N(t,\tau)$ of active particles generated by a seed at the origin
        \begin{equation}
           N(t) = \int d^d x\mbox{ } G_{\psi \tilde{\psi}}(t,\mx).
           \label{eq:scale_N}
        \end{equation}
  \item The mean square radius $R^2(t)$ of percolating particles, which 
        started from the origin at time $t=0$
        \begin{equation}
           R^2(t) = \frac{\int d^d\mx\mbox{ } \mx^2 G_{\psi \tilde{\psi}}(t,\mx)}
           {2d\int d^d\mx\mbox{ }  G_{\psi \tilde{\psi}}(t,\mx) }.
           \label{eq:scale_R}
        \end{equation}
  \item Survival probability $P(t)$ of an active cluster originating
        from a seed at the origin
        \begin{equation}
           P(t) = - \lim_{k\rightarrow\infty} \langle 
           \tilde{\psi}(-t,{\bm 0}) \eRM^{-k\int d^d\mx\mbox{ } \psi(0,\mx)}
           \rangle.
           \label{eq:scale_P}
        \end{equation}
\end{enumerate}

By straightforward analysis \cite{JanTau04} it can be shown that the scaling behavior of
these functions is given by the asymptotic relations
\begin{align}
   & R^2(t) \sim 
   t^{2/\Delta_\omega},
   \\
   & N(t) \sim t^{-(\gamma_\psi +  \gamma_{\tilde{\psi}})/\Delta_\omega},\\
   & P(t) \sim t^{-(d+\gamma_\psi +  \gamma_{\tilde{\psi}})/2\Delta_\omega} .
   \label{eq:scalingGF}
\end{align}

From the structure of anomalous dimensions 
(\ref{eq:gen_anom_charges}-\ref{eq:gen_anom_fields}) it is clear that the
resulting system of equations for FPs is quite complicated. Although
to some extent it is possible to obtain coordinates of the fixed points, the
eigenvalues of the matrix (\ref{eq:matrix}) pose a more severe technical problem.
Hence, in order to gain some physical insight into the 
structure of the model,  we divide overall analysis into special cases and
analyze them separately.
\subsection{Rapid change \label{subsec:rapid}}
First, we perform an analysis of the rapid-change limit of the model.
It is convenient \cite{Ant99,Ant00}
to introduce the new variables $g'_1$ and $w $ given by
\begin{equation}
   g_1' = \frac{g_1}{u_1}, \quad w=\frac{1}{u_1}.
   \label{eq:def_rapid}
\end{equation}
The rapid change limit then corresponds to fixed points with a coordinate $w^* = 0$.
Using the definition (\ref{eq:def_beta}) together with 
expressions (\ref{eq:beta_functions}) the beta-functions for the charges (\ref{eq:def_rapid})
are easily obtained 
\begin{align}
   & \beta_{g_1'} = g_1' (\eta - y +\gamma_D - 2\gamma_v) , \quad
   & &\beta_{w} = w(\eta - \gamma_D ).   
\end{align}
Analyzing the resulting system of equations  seven possible regimes
 can be found. Their coordinates are listed in Tab. \ref{tab:rchm} in 
Appendix \ref{app:fixed}. Due to the cumbersome form of the matrix
(\ref{eq:matrix}),  we were not able to determine all the corresponding eigenvalues in 
 an explicit form. In particular, for nontrivial fixed points (with non-zero coordinates 
 of $g_1',g_2$ and $u_2$) the resulting expressions
are of a quite unpleasant form. Nevertheless, using numerical software \cite{mathematica}
it is possible to obtain
all the necessary information about the fixed points' structure and in this way
the boundaries between the corresponding regimes have been obtained.
In the analysis it is advantageous to exploit additional constraints
following from the physical interpretation of the charges. For example,  
 $g_1'$ describes the density of kinetic energy of the velocity fluctuations,
 $g_2$ is equal to $\lambda^2$ and $a'$ will be later
 on introduced (see Appendix \ref{app:fixed}) as $(1-2a)^2$. Hence, it is clear that these
 parameters have to be non-negative real numbers. Fixed points that 
 violate this condition can be immediately discarded as non-physical.
 
 Out of seven possible fixed points, only 
four are IR stable: \fp{I}{1}, \fp{I}{2}, \fp{I}{5} and \fp{I}{6}. 
Thus,  only regimes which correspond to those points
could be in principle realized in real physical systems. As
expected \cite{Ant00},  the coordinates of these fixed points (see Tab.~\ref{tab:rchm}) and the scaling behavior 
of the Green functions (see Tab.~\ref{tab:exponents})
depend only on the parameter $\xi=y-\eta$.
In what follows, we restrict our discussion
only to them. 

\begin{figure}[h!]
  \includegraphics[width=7cm]{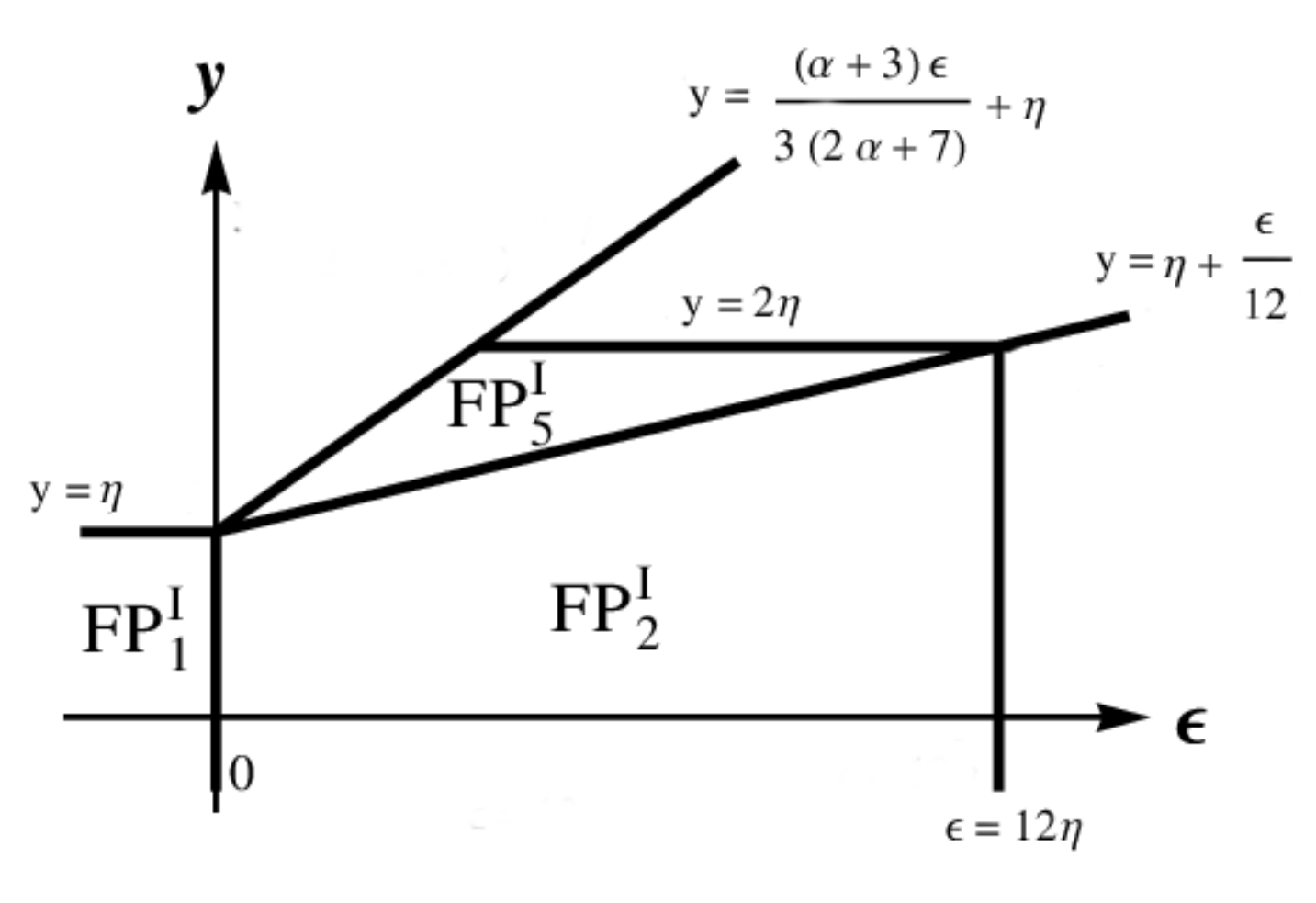}
  \caption{A qualitative sketch of the 
  regions of stability for the fixed points in the limit of the rapid-change model.
%
%
The borders between the regions are depicted with the bold lines.
%
%
  }
  \label{fig:stab_rch}
\end{figure}

The \fp{I}{1} represents the free (Gaussian) FP for which all interactions are
irrelevant and ordinary perturbation theory is applicable.
As expected, this regime is IR stable  in the region
\begin{equation}
   y < \eta,\quad \eta>0,\quad \eps < 0.
   \label{eq:FPI1_region}
\end{equation}
  The latter condition ensures that we are above the upper
 critical dimension $d_c=4$.
For  \fp{I}{2} the correlator of the velocity field is irrelevant and 
this point describes standard the DP universality class \cite{JanTau04} and is IR stable 
in the region
\begin{equation}
  \eps > 0,\quad \eps/12 + \eta > y,\quad \eps < 12\eta. 
  \label{eq:FPI2_region}
\end{equation} 
The remaining two fixed points constitute nontrivial regimes for which velocity
fluctuations as well as percolation interaction become relevant.
The \fp{I}{5} is IR stable in the region given by
\begin{equation}
 (\alpha+3)\eps > 3(2\alpha+7)(y-\eta),\quad 12(y-\eta)>\eps,\quad 2\eta>y.
 \label{eq:FPI5_region}
\end{equation}
The boundaries for \fp{I}{6} can be only computed by numerical calculations. 

\begin{figure}[h!]
  \includegraphics[width=6cm]{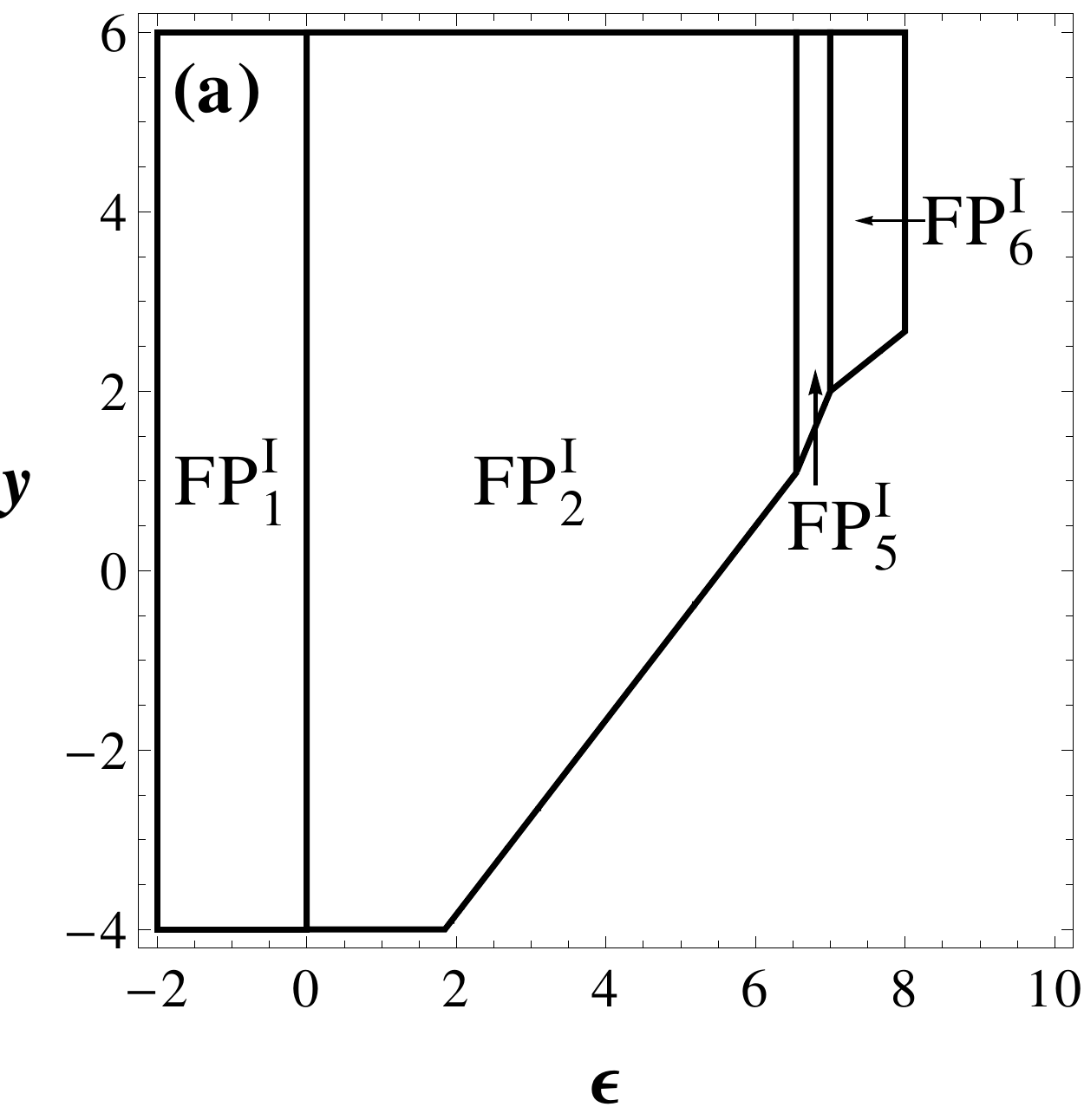}
  \includegraphics[width=6cm]{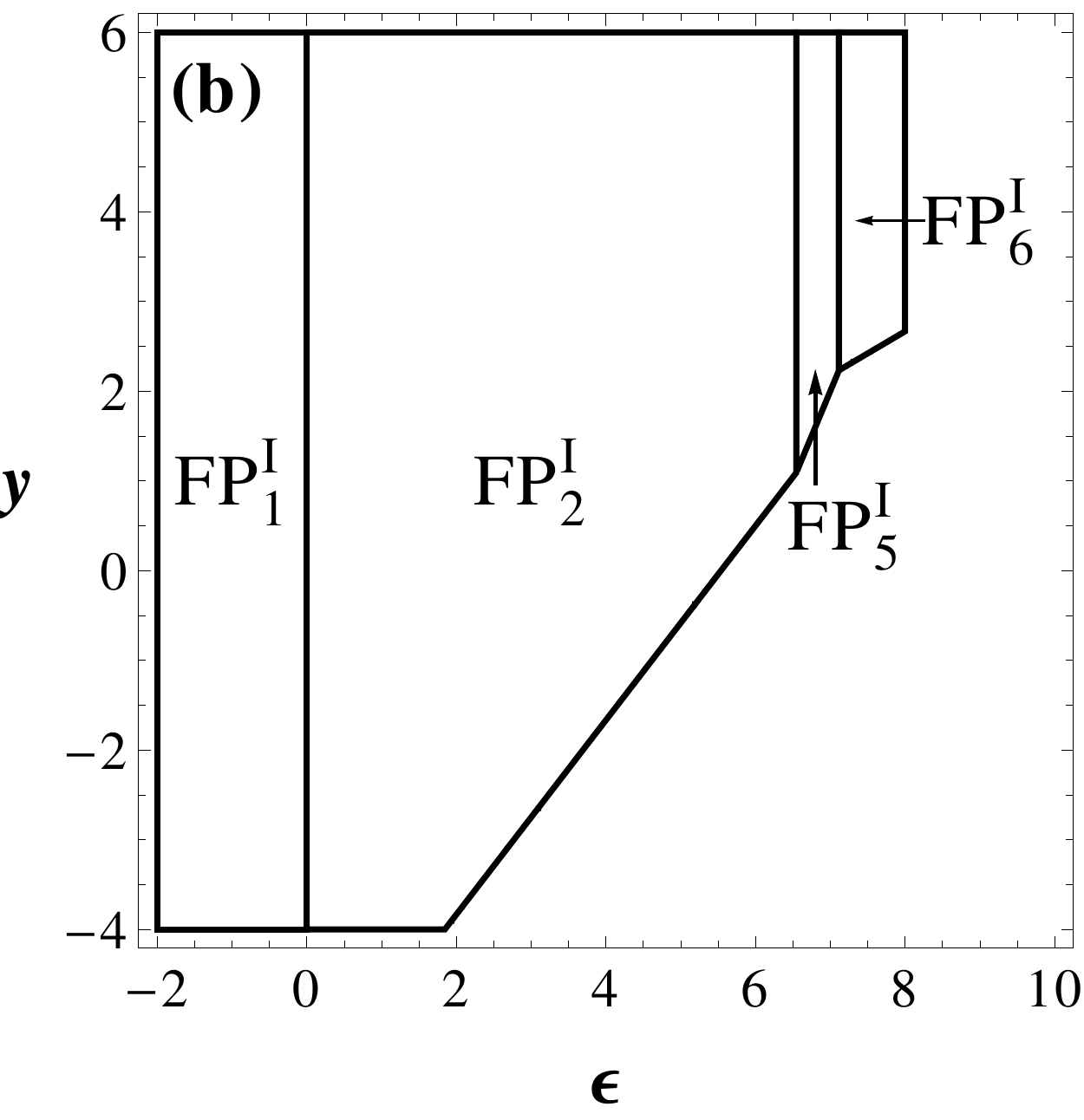}
  \includegraphics[width=6cm]{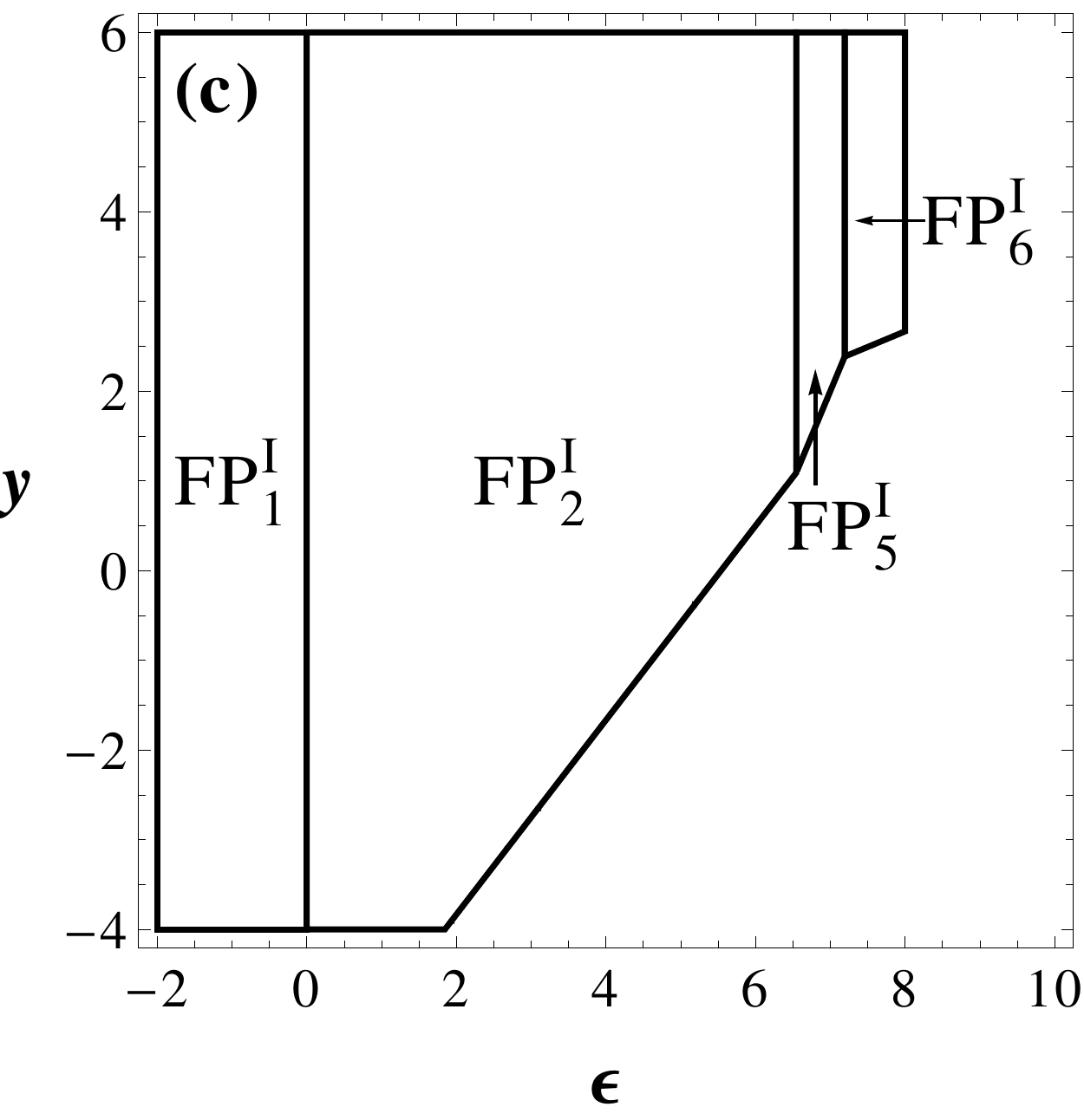}
  \caption{Fixed points' structure for the thermal noise situation 
	(\ref{eq:cond_thermal}). 
%
%
	From above to bottom the compressibility parameter $\alpha$ attains
	consecutively the values: (a) $\alpha=0$, (b) $\alpha=5$ and (c) 
	$\alpha=100$.	}
%
%

  \label{fig:thermal}	
\end{figure}

Using the information about the phase boundaries, a qualitative picture of the phase diagram
can be constructed. In Fig. \ref{fig:stab_rch} the situation in the plane $(\eps,y)$ is depicted.
We observe that compressibility affects 
 only the outer boundary of \fp{I}{5}. The larger value of $\alpha$ the larger area
of stability. We also observe that the realizability of the regime \fp{I}{5} 
crucially depends on the nonzero value of $\eta$.

The important subclass of the rapid-change  limit constitutes thermal velocity
fluctuations, which
are characterized by the quadratic dispersion law \cite{FNS77}. In our formulation this is achieved
by considering the following relation:
\begin{equation}
  \eta = 6 +y - \eps   
  \label{eq:cond_thermal}
\end{equation}
which follows directly from  expression (\ref{eq:rch_limit}).
The situation for increasing values of the parameter $\alpha$ is depicted in
Fig. \ref{fig:thermal}. We see that for physical space dimensions 
$d=3\mbox{ }(\eps=1)$ and $d=2\mbox{ }(\eps=2)$ the only stable regime is that of pure DP.
The nontrivial regimes \fp{I}{5} and \fp{I}{6} are realized only in the nonphysical
region for large values of $\eps$. This numerical result confirms our
previous expectations \cite{AntKap08,AntKap10}. 
It was pointed out \cite{HH00,HHL13} that
genuine thermal fluctuations could change IR stability of the given universality class. 
However, this is not realized for the percolation process.
\subsection{Regime of frozen velocity field \label{subsec:frozen}}

According to  equation (\ref{eq:fvf_limit}),  the regime of  the frozen
velocity field
corresponds to the constraint $u_{1}^*=0$.
\begin{figure}[h!]
  \includegraphics[width=6cm]{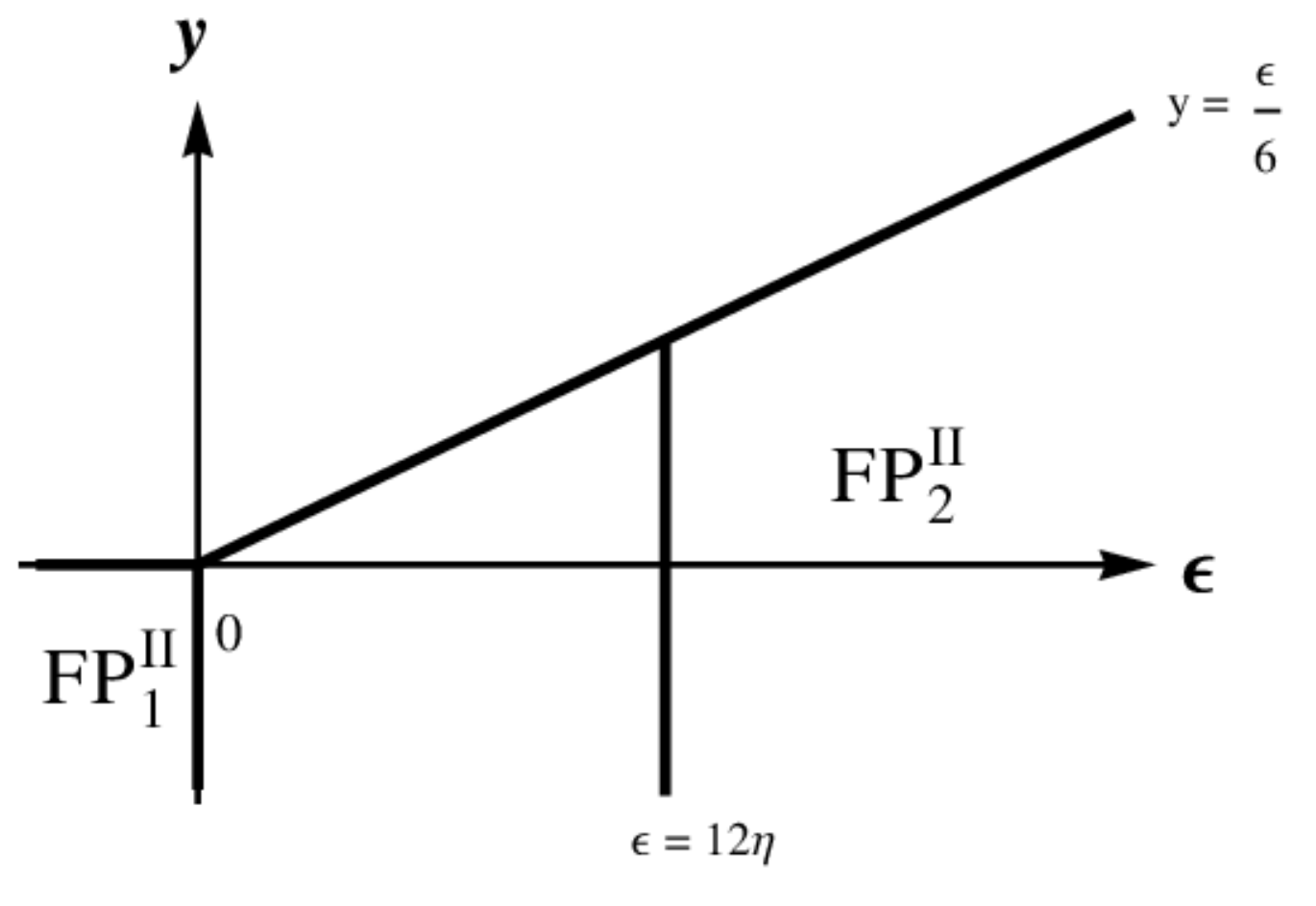}
  \caption{Regions of stability for the fixed points in the limit of  the  frozen velocity field.
%
%
  The borders between the regions are depicted with the bold lines.
%
%
  }
  \label{fig:stab_fvf}
\end{figure}
 Using a general form of anomalous dimensions (\ref{eq:gen_anom_charges}) with
 the given constraint on $u_1$ eight possible fixed points are obtained.
Their coordinates are listed in Tab. \ref{tab:fvf} in Appendix \ref{app:fixed}. However, only three
of them (\fp{II}{1}, \fp{II}{2} and \fp{II}{7}) could be physically realized (IR stable).
 
\begin{figure}[h!]
  \includegraphics[width=6cm]{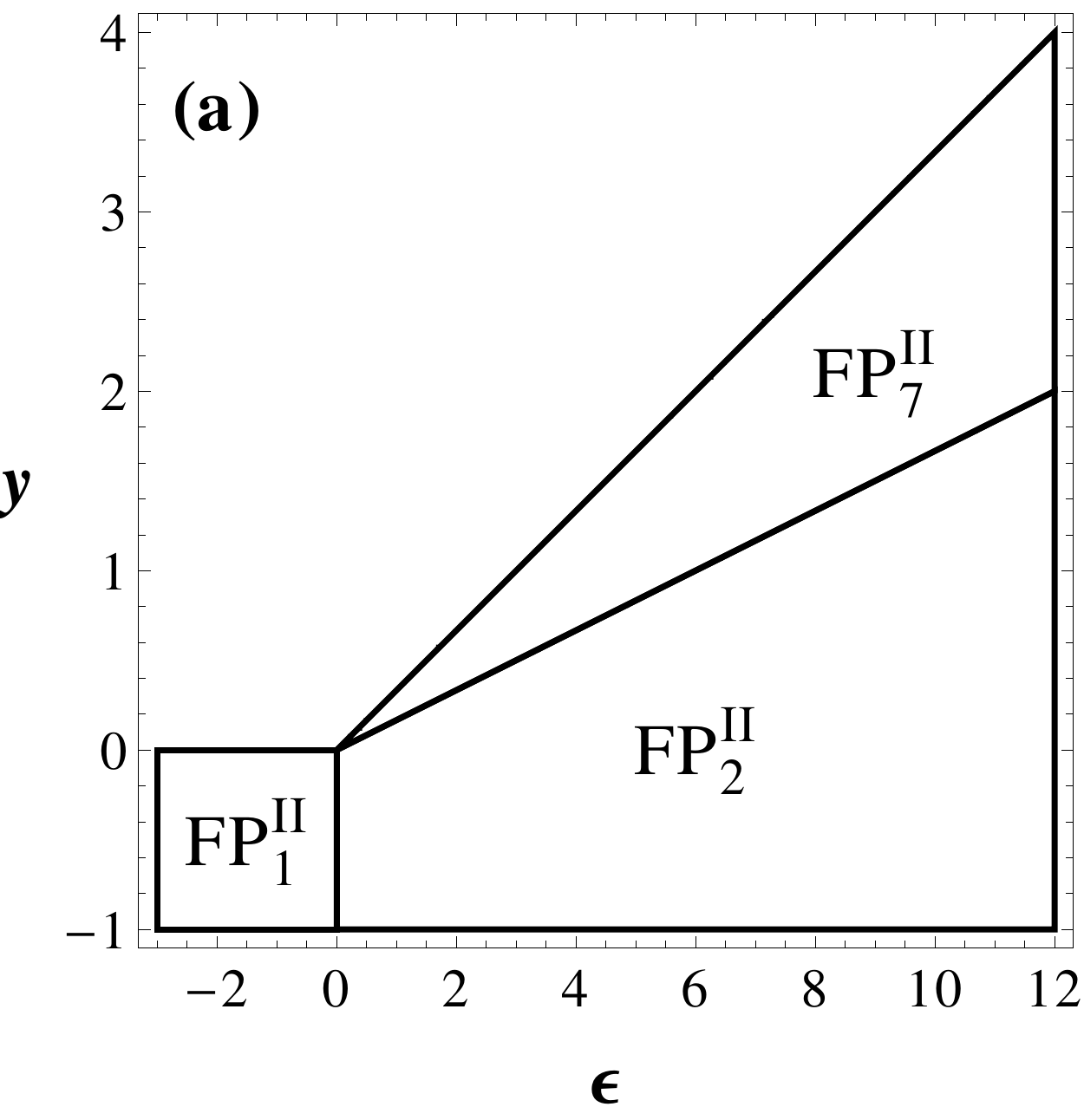}
  \includegraphics[width=6cm]{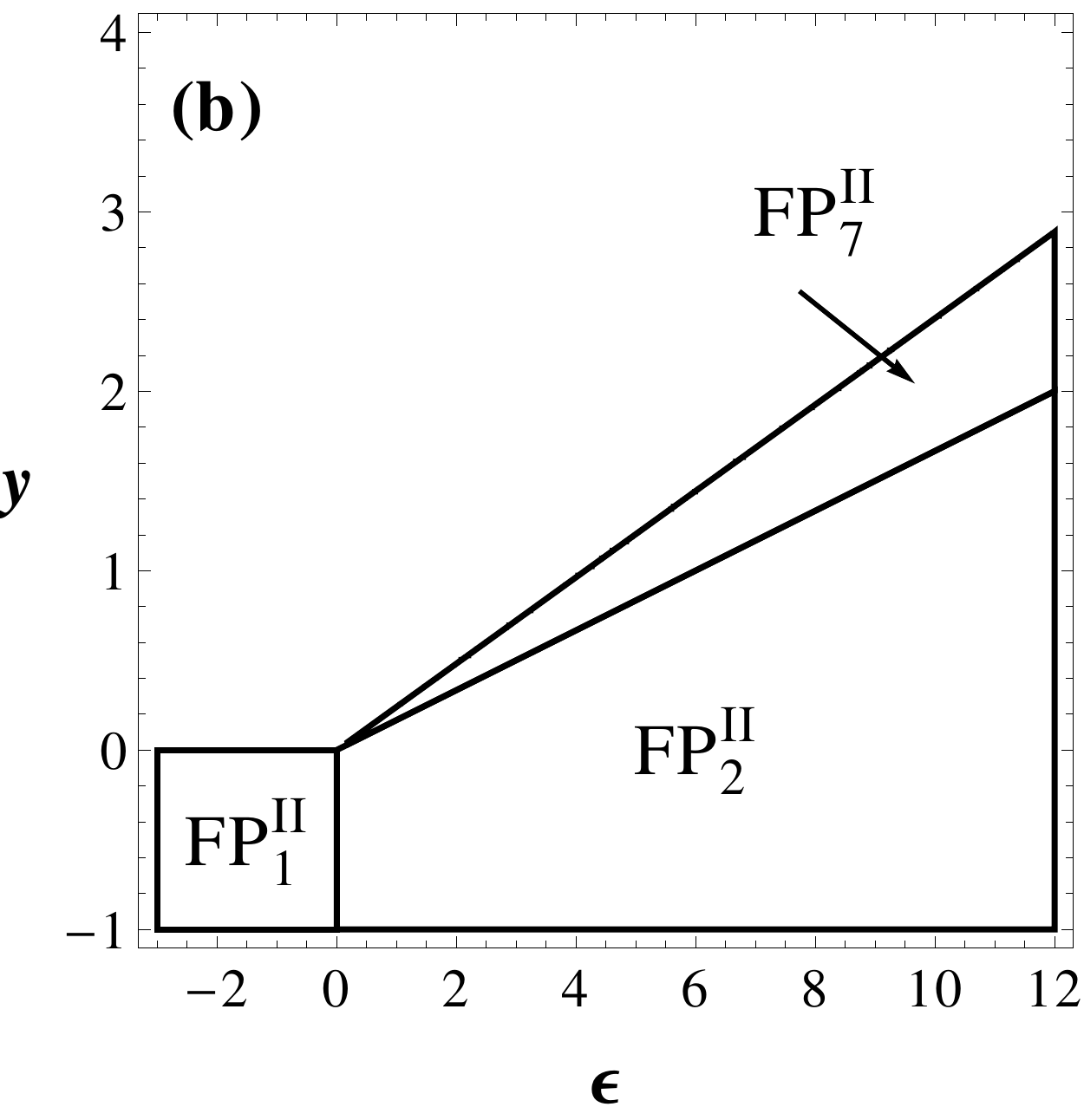}
  \includegraphics[width=6cm]{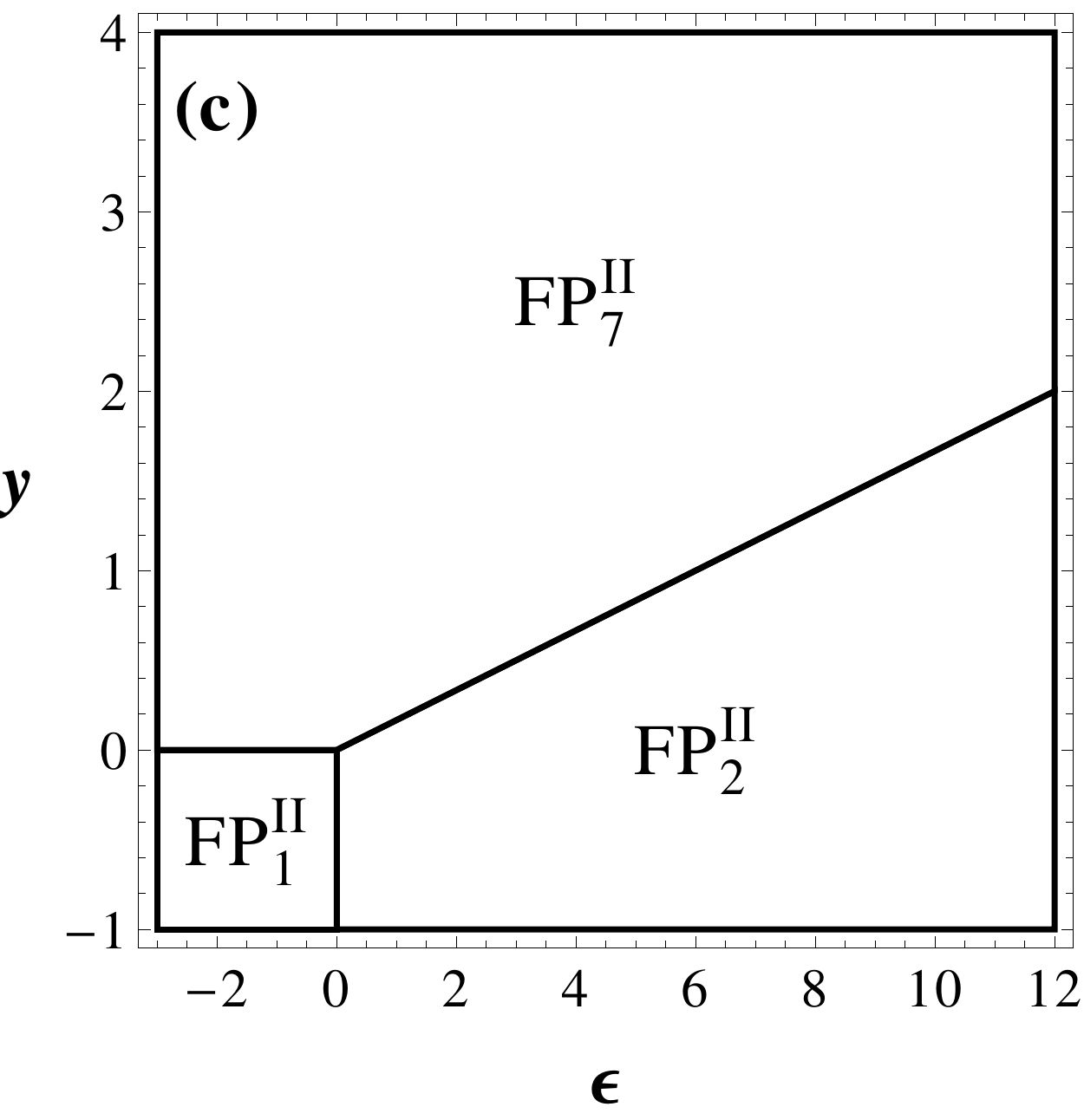}
  \caption{Fixed points' structure for frozen velocity case
        with $\eta=0$.
%
%
	From above to bottom the compressibility parameter $\alpha$ attains
	consecutively the values: (a) $\alpha=0$, |(b) $\alpha=3.5$ and (c) $\alpha=8$.}
%
%
	\label{fig:FVF_struct}	
\end{figure}
The fixed point \fp{II}{1} describes the free (Gaussian) theory. It is stable in the region
\begin{equation}
   y<0,\quad \eps < 0, \quad \eta < 0.
   \label{eq:free_frozen} 
\end{equation}

For  \fp{II}{2}
the velocity field is asymptotically irrelevant and the only 
relevant interaction is due to the percolation process itself. This regime is stable in
 the region
\begin{equation}
   \eps > 6y, \quad \eps > 0,\quad \eps > 12\eta.
   \label{eq:DP_frozen}
\end{equation}

On the other hand,  \fp{II}{7} represents a truly nontrivial regime for which both velocity and
percolation are relevant. The regions of stability for the \fp{II}{1} and \fp{II}{2} 
are depicted in Fig. \ref{fig:stab_fvf}. Since for these two points
the velocity field could be effectively neglected, the trivial observation is that
these boundaries do not depend on the value of the parameter $\alpha$. The stability region
of \fp{II}{7} can be computed only numerically. 

In order to study the  influence of compressibility on the stability in the nontrivial regime
\fp{II}{7},  we have studied situation for $\eta=0$. For other values of $\eta$ the situation
remains qualitatively the same. The situation for increasing values of $\alpha$ is depicted
 in Fig. \ref{fig:FVF_struct}. We observe that for $\alpha=0$ there is a region
of stability for \fp{II}{7}, which shrinks for the immediate value $\alpha=3.5$ to a smaller
area. Numerical
analysis shows that this shrinking continues well down to the value $\alpha=6$. A further increase
 of $\alpha$ leads to a substantially larger region of stability for the given
FP. Already for $\alpha=8$ this region covers all the rest of the $(y,\eps)$ plane.
The compressibility thus changes profoundly a simple picture expected from an incompressible case.
 Altogether the advection process becomes
more efficient due to the combined effects of  compressibility and the nonlinear terms.
\subsection{Turbulent advection \label{subsec:turbulent}}
In the last part we focus on a special case of the turbulent advection.
Our main aim is to determine whether Kolmogorov regime \cite{Frisch}, which
corresponds to the choice $y=2\eta=8/3$, could  lead to  a new
nontrivial regime for the percolation process.
In this section,  the parameter $\eta$ is always considered to attain its Kolmogorov value, $4/3$.
For a better visualization we present two-dimensional regions of stability
in the plane $(\eps,y)$ for different values of the parameter $\alpha$.
\begin{figure}[h!]
  \includegraphics[width=6cm]{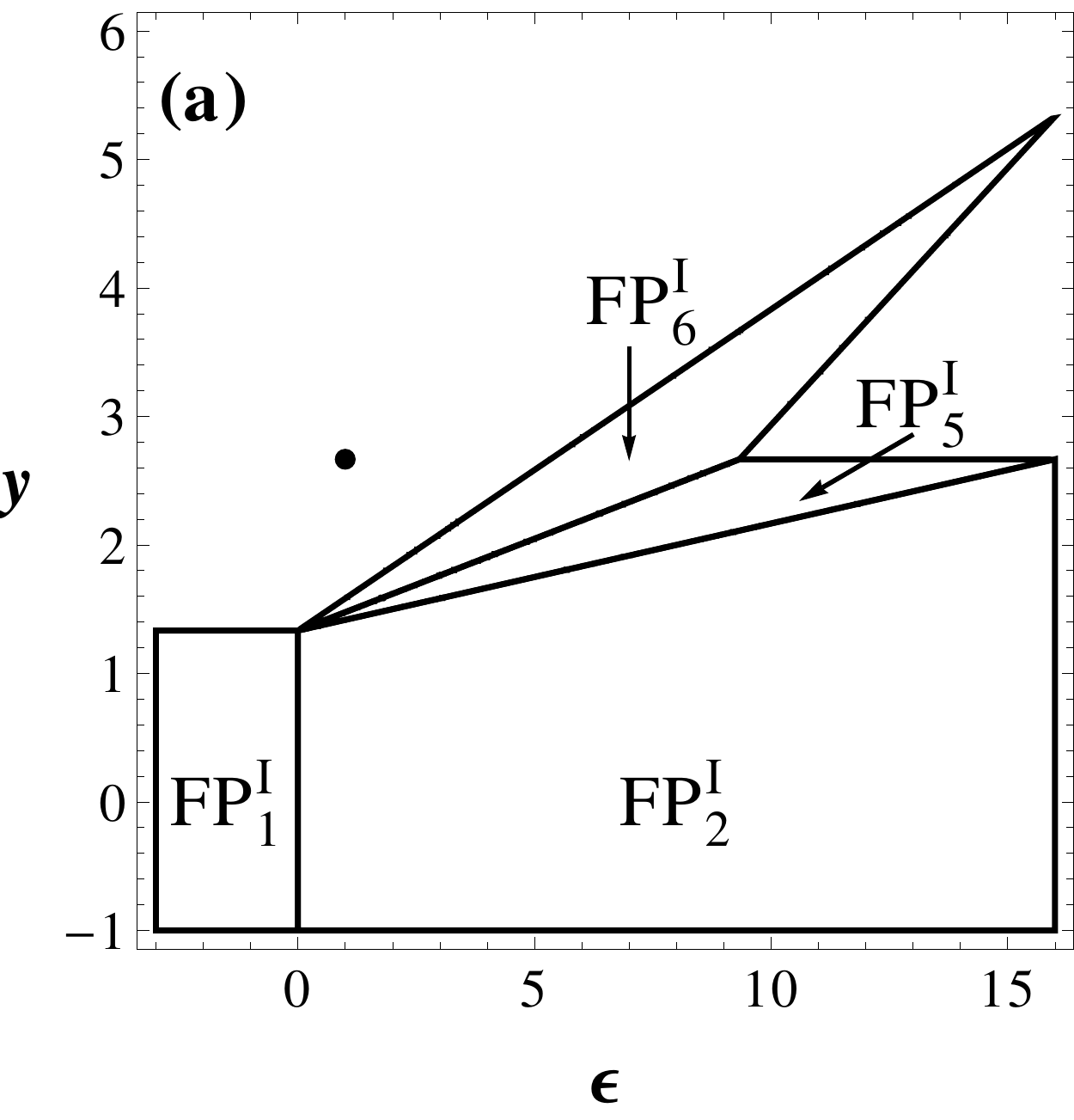} 
  \includegraphics[width=6cm]{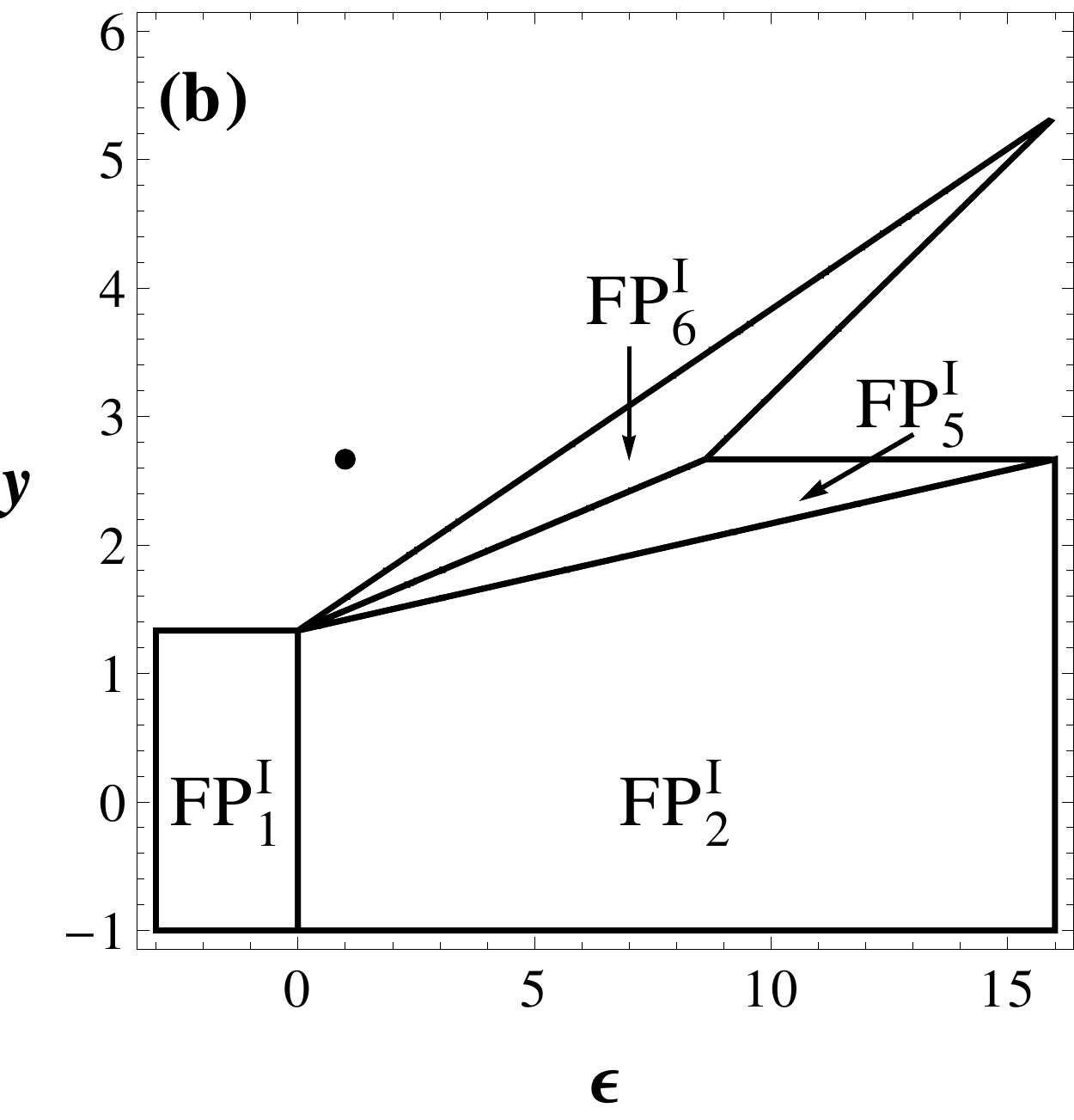}
  \includegraphics[width=6cm]{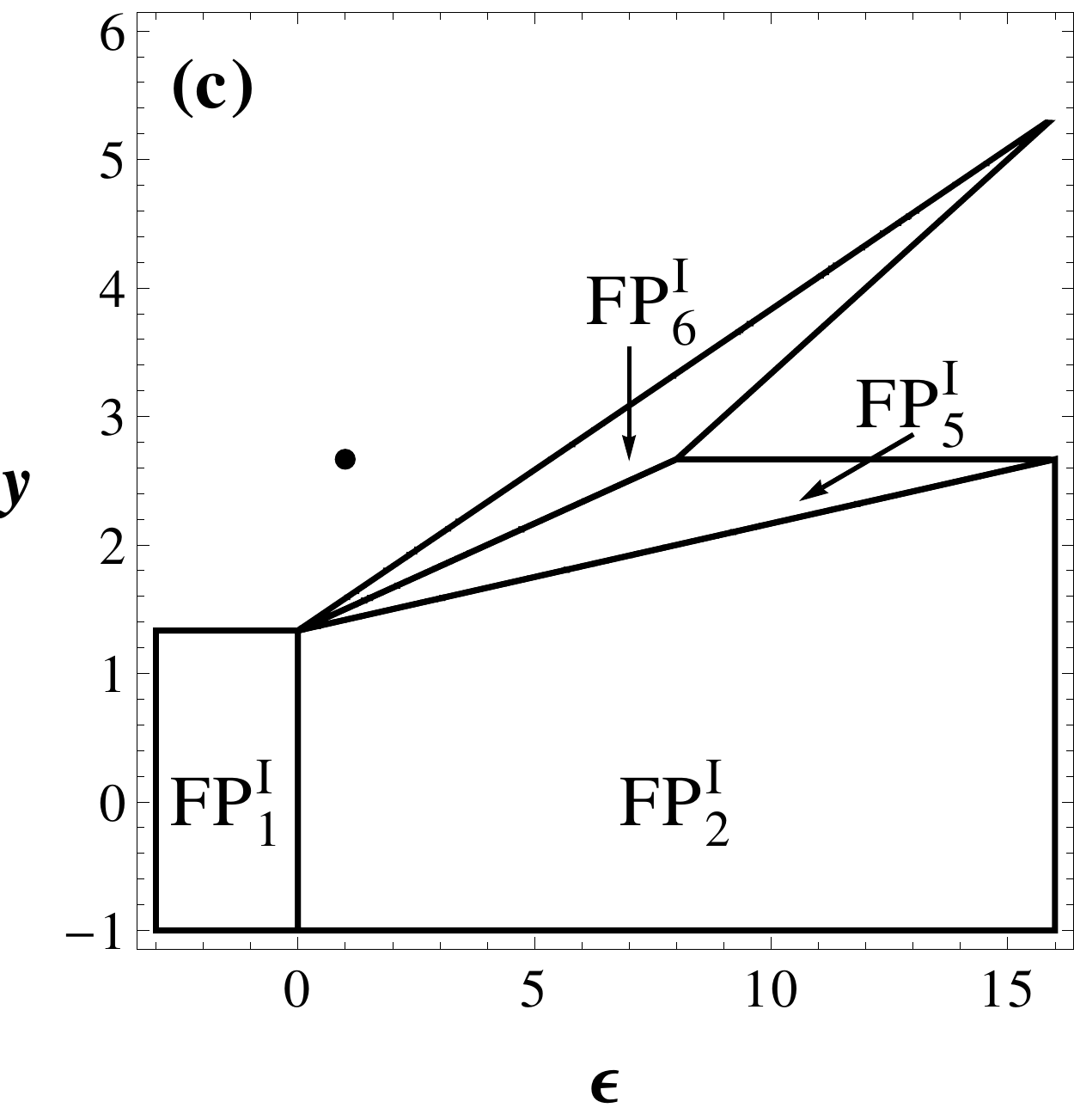}  
  \caption{  
%
%
  Fixed points' structure for rapid change model with $\eta=4/3$. 
	From above to bottom the compressibility parameter $\alpha$ attains
	consecutively the values: (a) $\alpha=0$, (b) $\alpha=5$ and (c) $\alpha=\infty$. 
	The dot denotes the coordinates of the
	three-dimensional Kolmogorov regime.}
%
%
  \label{fig:rapid}	
\end{figure}

First of all, we reanalyze the situation for the rapid-change model. The result
is depicted in Fig.~\ref{fig:rapid}. It is clearly visible that for this case
 a realistic turbulent scenario ($\varepsilon=1$ or $\varepsilon=2$) falls out of the 
possible stable regions. This result is expected because the rapid-change model with
vanishing time-correlations could not properly describe well-known
turbulent properties \cite{Frisch,Monin}. We also observe that compressibility
mainly affects the boundaries between the regions \fp{I}{5} and \fp{I}{6}. However, this
happens mainly in the nonphysical region.

\begin{figure}[h!]
  \includegraphics[width=6cm]{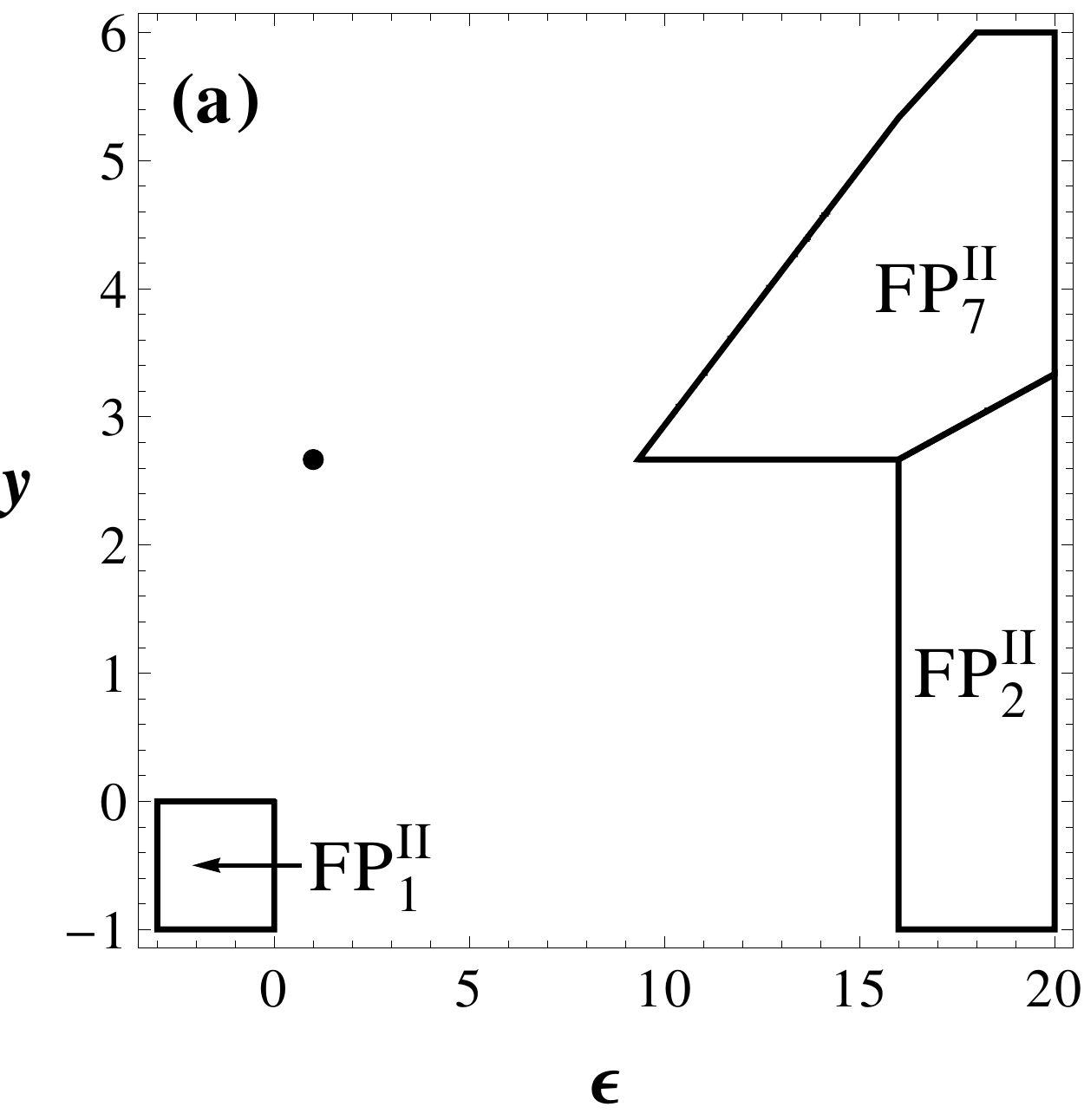}
  \includegraphics[width=6cm]{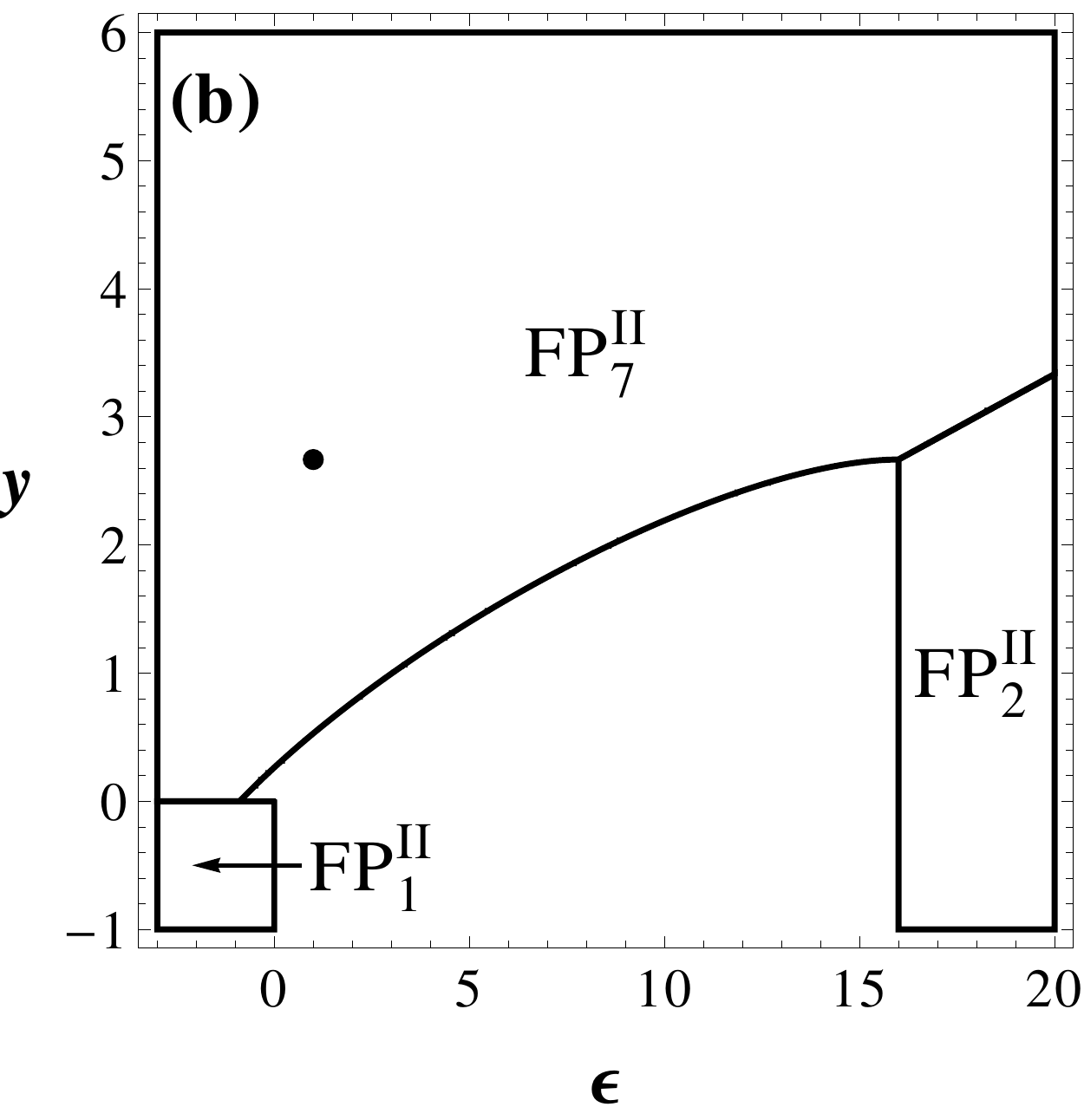}
  \includegraphics[width=6cm]{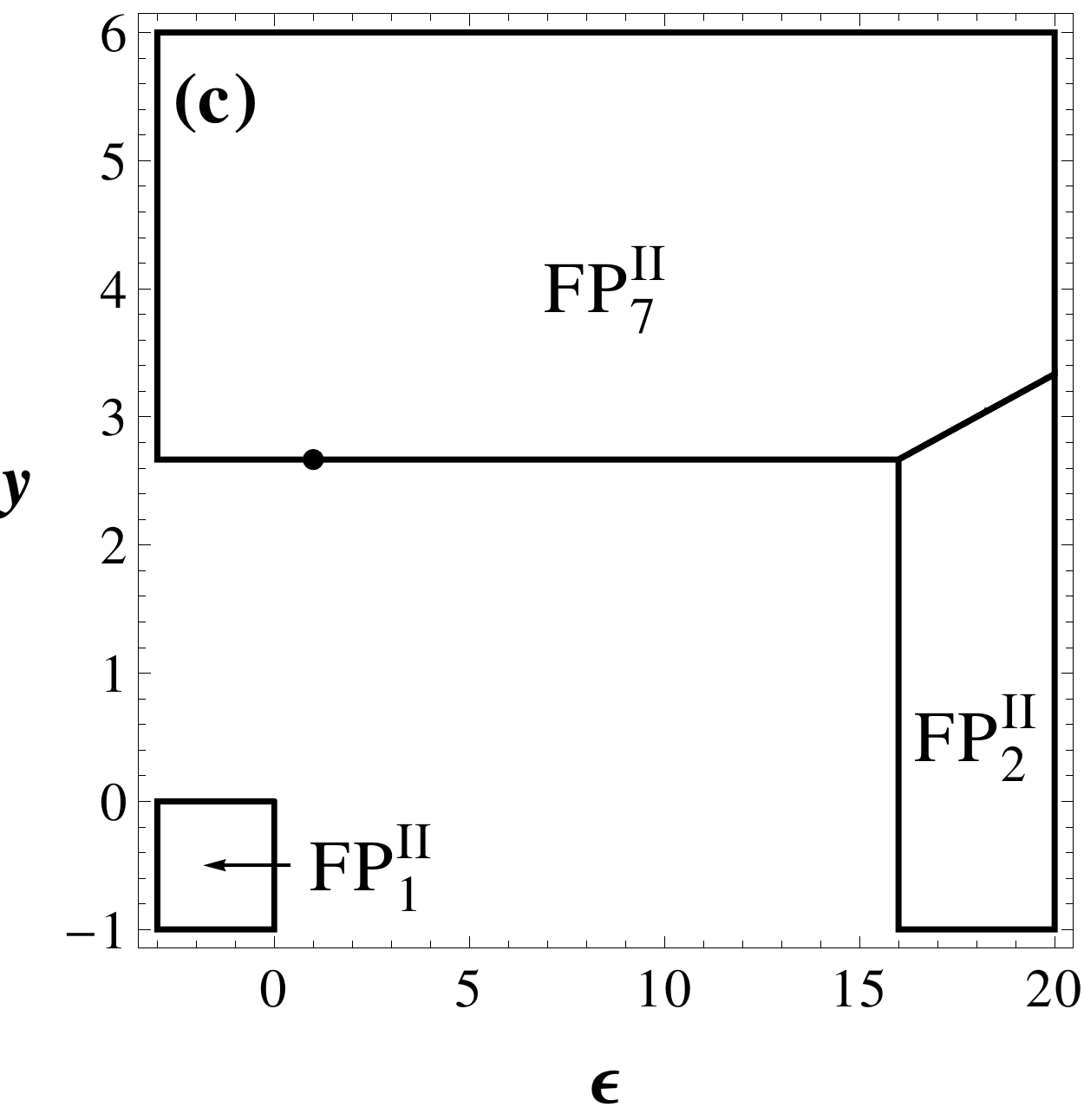}
  \caption{
%
%
  Fixed points' structure for  the frozen velocity case
        with $\eta=4/3$. From above to bottom the compressibility parameter $\alpha$ attains
	consecutively the values: (a) $\alpha=0$, (b) $\alpha=8$ and (c) $\alpha=\infty$.
	The dot denotes the coordinates of  the
	three-dimensional Kolmogorov regime.
  }
%
%
  \label{fig:frozen}  
\end{figure}

Next, we turn our attention to a
similar analysis for the frozen velocity field. The corresponding stability
regions are depicted in Fig. \ref{fig:frozen}.
Here we see that the situation is more complex. 
The regime \fp{II}{2} is situated in the non-physical region and could not be realized. 
For small values of  the parameter $\alpha$
the Kolmogorov regime (depicted by a point) does not belong to the frozen velocity limit. 
However,
from a special value $\alpha = 6$ up to $\alpha\rightarrow\infty$ the Kolmogorov
regime belongs to the frozen velocity limit. Note that the bottom line for the region of
stability of \fp{II}{7} is exactly given by $y=4/3$. We observe that compressibility affects
mainly the boundary of the nontrivial region. We conclude that the presence of 
compressibility has  a stabilizing effect on the regimes where nonlinearities are relevant.

Finally, we look carefully at the nontrivial regime, which means that no
special requirements were laid upon the parameter $u_1$. As obtaining of 
analytical results proves to be too difficult, we have analyzed numerically the differential
equations for the RG flow (\ref{eq:invariant_chrg}). 
We found that the behavior of the  RG flows is as follows.
There exists a borderline in the plane $(\varepsilon,\alpha_c)$ given approximately by the expression
\begin{equation}
  \alpha_c = -12.131\varepsilon + 117.165.
  \label{eq:alpha_crit}
\end{equation}
Below $\alpha_c$,
only the frozen velocity regime corresponding to \fp{II}{7} is stable. 
Above $\alpha_c$, three fixed points \fp{II}{7}, \fp{III}{1} and \fp{III}{2} are observed. Whereas
two of them (\fp{II}{7} and \fp{III}{1}) 
are IR stable, the remaining one \fp{III}{2} is unstable in the IR regime. Again one of the stable FPs
 corresponds to \fp{II}{7}, but the new FP is a regime with finite correlation time. 
 For the reference the coordinates of these two points for the value $\alpha=110$
 are given in Tabs. \ref{tab:nontrivial1} and \ref{tab:nontrivial2} in Appendix \ref{app:fixed}.
 Since all free parameters $(\eps,\eta,y,\alpha)$ are the same for both points, 
 which of the two points
 will be realized depends on the initial values of the bare parameters.
A similar situation is observed for the stochastic magnetohydrodynamic
turbulence \cite{MHD01}, where the crucial role is played by a forcing decay-parameter.
\begin{figure}[h!]
  \includegraphics[width=6cm]{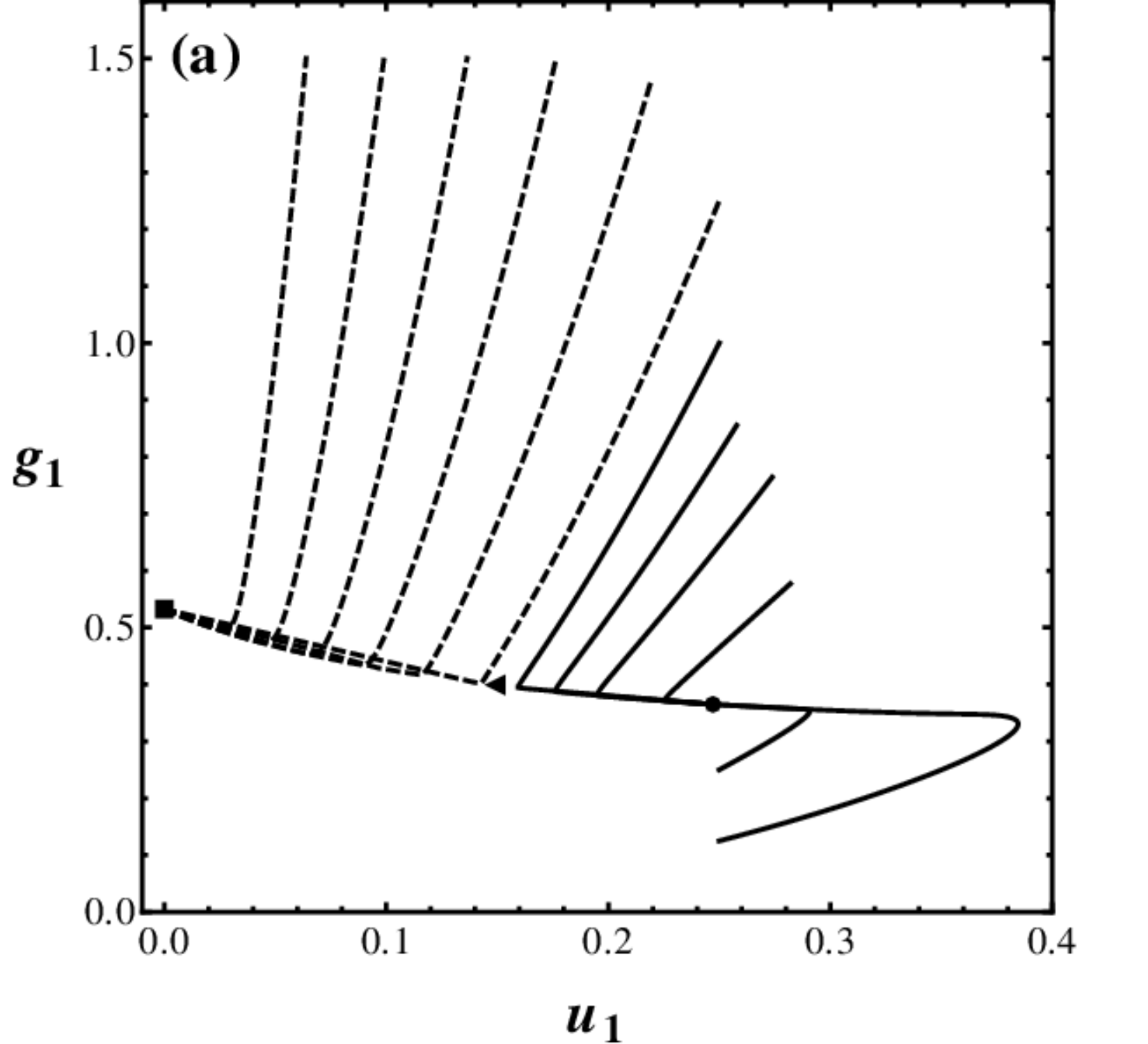} 
   \includegraphics[width=6cm]{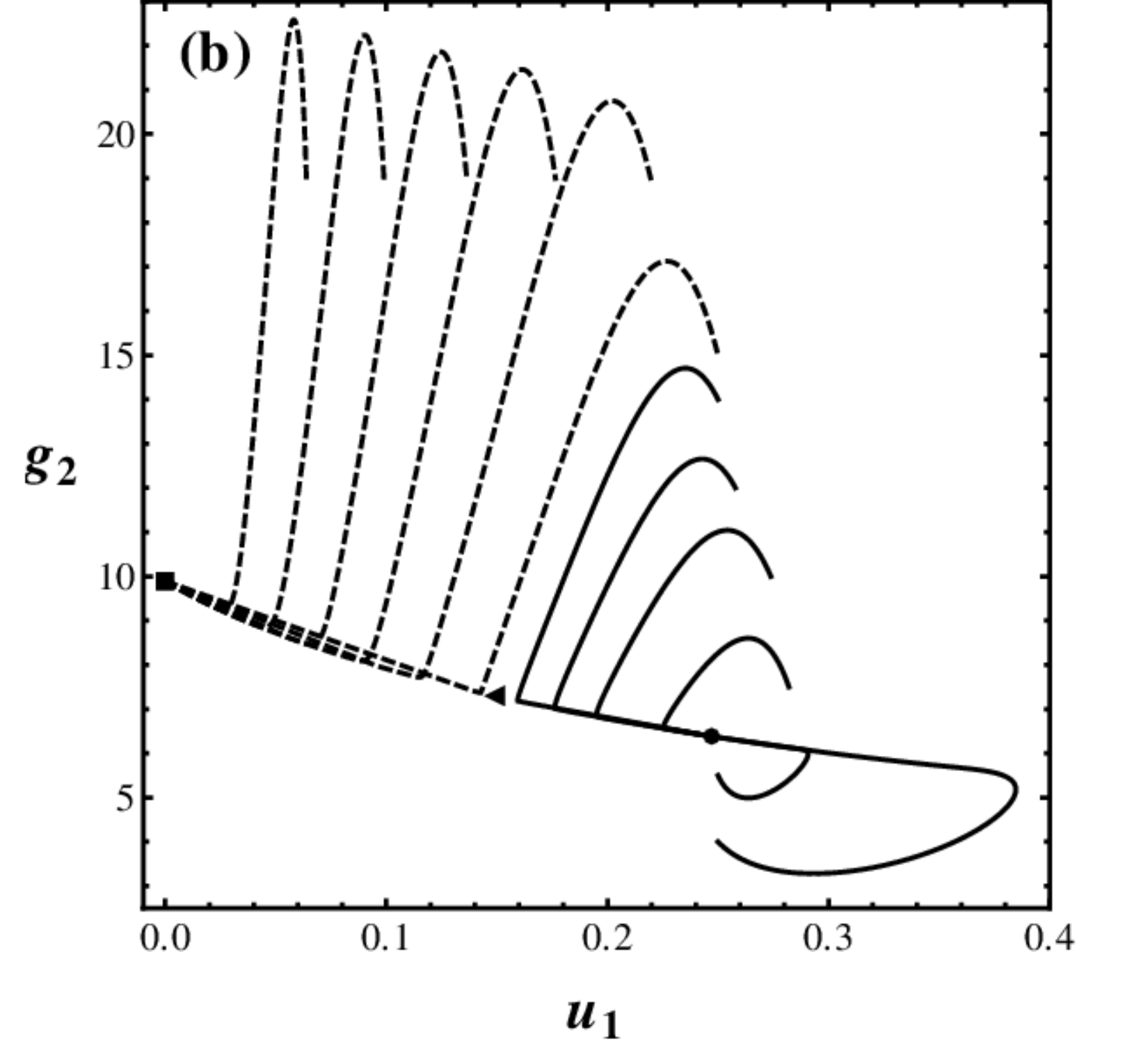}  
  \caption{
%
%
    Demonstration of the
    RG trajectories flows' in:  (a) the plane $(g_1,u_1)$ and (b) the plane $(g_2,u_1)$ for three dimensional
    ($\eps=1$) turbulent advection with
    $\alpha=110$. The square $\blacksquare$  denotes frozen velocity regime \fp{II}{7}, 
    triangle $\blacktriangleleft$ corresponds to the unstable regime \fp{III}{2} and
    circle $\bullet$ to the nontrivial
    regime \fp{III}{1} for which the time correlations are relevant. Dashed lines
    corresponds to the chosen flows to the point \fp{II}{7}, whereas
    the full lines to the flows to the other stable point \fp{III}{1}.
    \label{fig:flow2D}  
  }
%
%
\end{figure}  

For illustration purposes the projections of the RG flow onto the planes $(g_1,u_1)$
and $(g_2,u_1)$ are
depicted in Fig. \ref{fig:flow2D}. The two stable points are clearly separated
by the unstable one.  

\section{Conclusions\label{sec:concl}}
In this paper,  we have studied an effect of compressibility on the paradigmatic model
of percolation spreading. The coarse grained model of percolation with
inclusion of the  advecting velocity field can be reformulated as the multiplicatively
renormalizable field theoretic model. 

We have found that depending on the values of a spatial dimension $d=4-\eps$, scaling
exponents $y$ and $\eta$, describing statistics of velocity fluctuations and
a degree of compressibility $\alpha$, the model exhibits $8$ distinct universality classes.
Some of them are already well-known: the Gaussian (free) fixed point,
 a directed percolation without advection and a passive scalar advection. The remaining
 points correspond to new universality classes, for which an interplay
between advection and percolation is relevant.

It also has to be kept in mind that  only relatively small values of
$\alpha$ are allowed ($\alpha \ll 1$) in our model. It corresponds to small fluctuations of the density 
$\rho$, what is tacitly supposed in our investigation. In other words, it is assumed that the
stochastic component of the velocity field of the fluid is much smaller than the velocity
of the sound in the system (the Mach number $\mathrm{Ma} \ll 1$).
Hence our results must be taken
with a grain of salt. Nevertheless we believe that a qualitative picture for
large values of compressibility should remain the same.
In order to properly
describe effects of strong compressibility and to better understand non-universal
effects for turbulent advection one should proceed one step further
and employ a more sophisticated model for velocity fluctuations, e.g. one
introduced in the works \cite{Volna96,ADU97} and later used for the passive advection problem
 in \cite{AntKos14}. However, such a model goes
well beyond the aim of this paper and its detailed study is left for a  future study.

\subsection*{Acknowledgments}
The work was supported by VEGA grant No. $1/0222/13$ 
 of the Ministry of Education, Science, Research and Sport of the Slovak Republic. 
 N.V.A. and A.S.K. acknowledge Saint Peterburg State University for Research 
 Grant No. 11.38.185.2014. A.S.K. was also supported by the grant 16-32-00086 provided
 by the Russian Foundation for Basic Research.
\appendix
  \section{Calculation of the Renormalization Constants \label{app:const}}  
In this appendix, we describe in detail how the renormalization constants
 are computed.

Though the bare action (\ref{eq:bare_act}) contains a lot of terms, the number
of divergent Feynman graphs is quite low to the first order of 
perturbation theory. Their analysis is to some extent simplified by  two facts:
\begin{enumerate}
 \item Integral of a power of internal momenta is zero in the dimensional regularization. 
       Hence, the tadpole diagrams are discarded.
 \item Closed circuits of propagators $\tilde{\psi}\psi$ vanish identically, which 
       is a consequence of the {It\^o} time discretization \cite{Gardiner,Kampen}, 
       which  we consider here.	  
\end{enumerate}
 For the two-point Green functions $\Gamma_{\tilde{\psi}\psi}$, the following
Dyson equation can be written
\begin{align}
  \Gamma_{\tilde{\psi}\psi}
  & = i\omega Z_1 - D p^2 Z_2
  -D\tau Z_3 + 
  \raisebox{-0.5ex}{
  \includegraphics[width=2.5truecm]{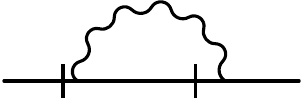}} \nonumber \\
  & + 
  \frac{1}{2}
  \raisebox{-.50ex}{
  \includegraphics[width=2.5truecm]{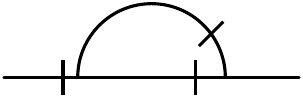} }.  
  \label{eq:exp_pp}
\end{align}
Up to the one-loop order the perturbation expansions for 
the vertex functions read consequently
\begin{align}
  \Gamma_{\tilde{\psi} {\psi} \mv}
  & =  
  -ip_j Z_4 - ia q_j Z_5 +
  \raisebox{-5.25ex}{  
   \includegraphics[width=2.truecm]{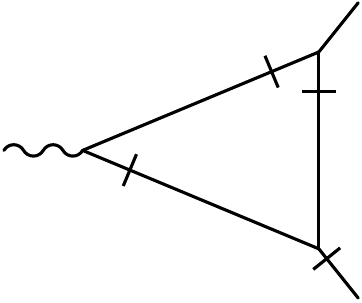}} 
   \nonumber \\
   & + 
  \raisebox{-5.25ex}{  \includegraphics[width=2.truecm]{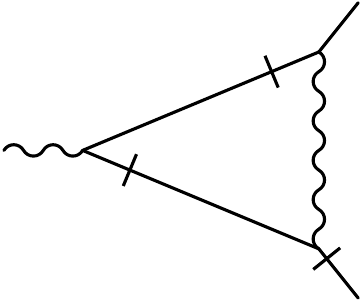}} 
   + \raisebox{-5.25ex}{  \includegraphics[width=2.5truecm]{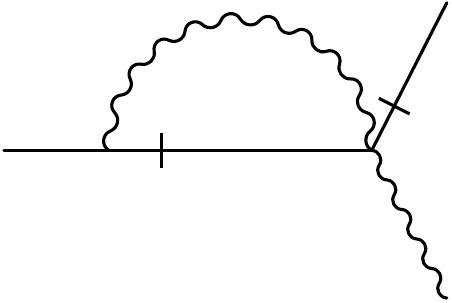}}
   \nonumber \\  
   & +
  \raisebox{-5.25ex}{  \includegraphics[width=2.5truecm]{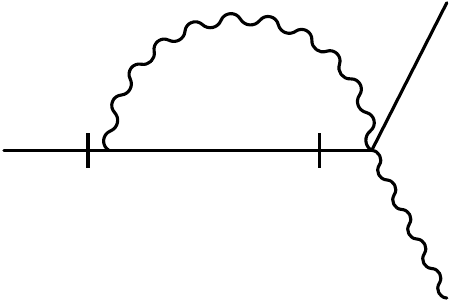}}.  
  \label{eq:exp_ppv}
\end{align}
\begin{align}
  \Gamma_{\tilde{\psi}\tilde{\psi} \psi}
  & = D\lambda Z_6 +
  \raisebox{-5.ex}{  \includegraphics[width=2.truecm]{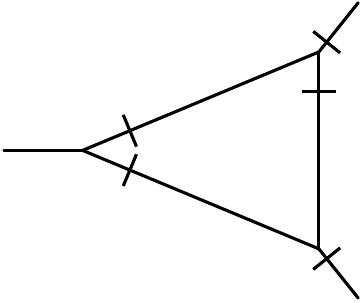}} 
   + 
  \frac{1}{2}
  \raisebox{-5.ex}{  \includegraphics[width=2.truecm]{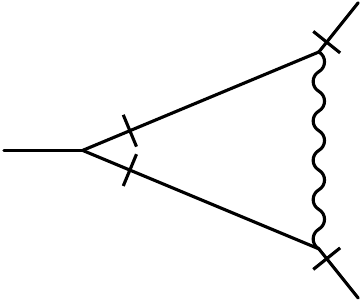}} 
   \nonumber \\
   & + 
  \raisebox{-5.ex}{  \includegraphics[width=2.truecm]{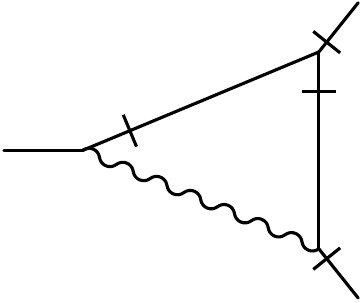}}.  
  \label{eq:exp_ppp1}
\end{align}
\begin{align}
  \Gamma_{\tilde{\psi}{\psi} \psi}
  & =  -D\lambda Z_7 +
  \raisebox{-5.ex}{  \includegraphics[width=2.truecm]{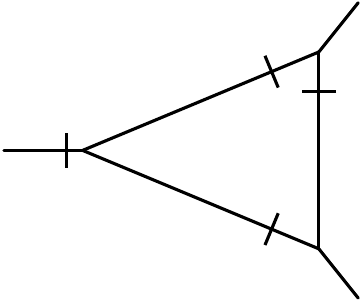}} 
   + 
  \frac{1}{2}
  \raisebox{-5.ex}{  \includegraphics[width=2truecm]{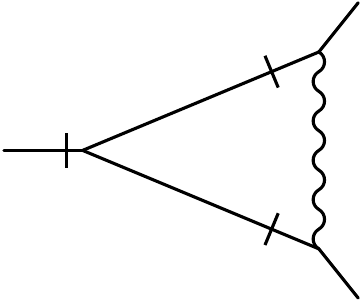}} 
   \nonumber \\
   & +
  \raisebox{-5.ex}{  \includegraphics[width=2truecm]{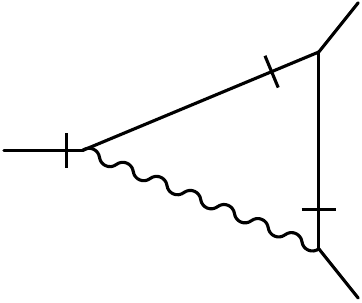}}  .
  \label{eq:exp_ppp2}
\end{align}
\begin{align}
  \Gamma_{\tilde{\psi}{\psi} \mv \mv}
  & =  \frac{u_{2}}{D}\delta_{ij} Z_8 +
  \raisebox{-5.25ex}{  
  \includegraphics[width=1.75truecm]{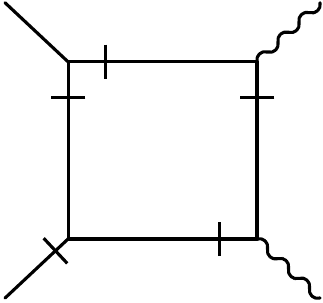}
  } 
   +   
  \raisebox{-5.ex}{  \includegraphics[width=1.75truecm]{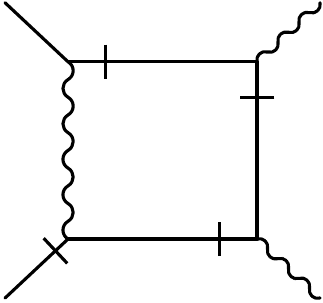}} 
   \nonumber \\
   & + \frac{1}{2}
  \raisebox{-5.ex}{  \includegraphics[width=1.75truecm]{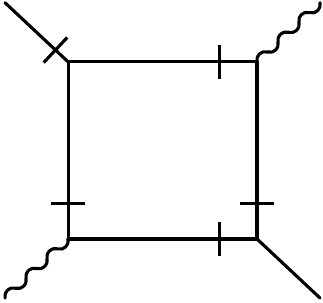}}  
  + \raisebox{-5.ex}{  \includegraphics[width=2.25truecm]{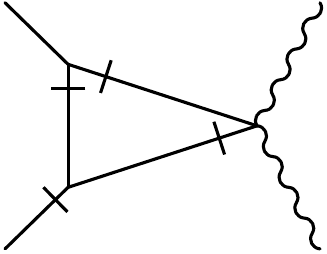}}  
  \nonumber \\
  & + 
  \raisebox{-5.25ex}{  \includegraphics[width=2.25truecm]{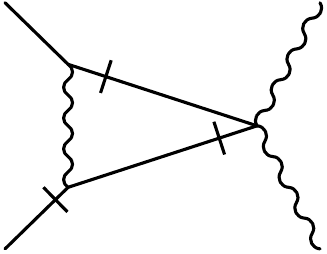}}  
  + \raisebox{-5.25ex}{  \includegraphics[width=2.25truecm]{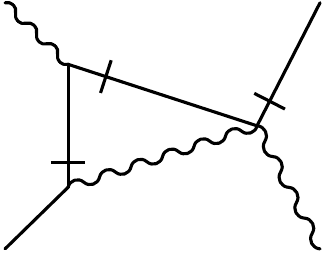}}  
  \nonumber \\
  & + 
  \raisebox{-5.25ex}{  \includegraphics[width=2.25truecm]{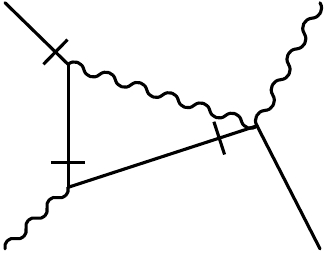}}  
  + \raisebox{-5.25ex}{  \includegraphics[width=2.25truecm]{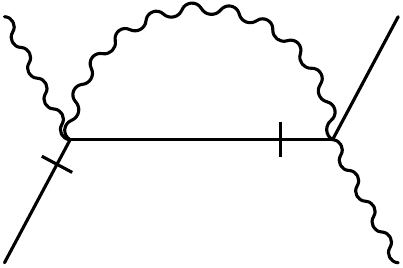}}   .
  \label{eq:exp_ppvv}
\end{align}
In these equations, we have explicitly given the symmetry coefficients \cite{Vasiliev}
of the corresponding diagrams. The numerical contributions arising from
variational derivatives with respect to the external fields are included
in the contribution of a given graph.
Note that in the language of Feynman graphs 
the need for the term $\propto \tilde{\psi}\psi\mv^2$ can be traced out
to the presence of the second Feynman graph in (\ref{eq:exp_ppvv}), which does not
vanish due to finite correlation time property of the velocity propagator 
 (\ref{eq:vel_correl}).

The computation of the diverging parts of the Feynman graphs
follows the standard methods
of dimensional regularization \cite{Amit,Vasiliev} and the 1-loop results
are
\begin{align}
   \label{eq:RG_inverse1}
   Z_1 & =  1 + \frac{g_1\alpha a(1-a)}{(1+u_1)^2y} + \frac{g_2}{4\eps},
   \nonumber  \\   
   Z_2 & =  1 - \frac{g_1}{4(1+u_1)y}\biggl[ 3+ 
     \alpha \NB{\frac{u_1-1}{u_1+1}-\frac{4a(1-a)u_1}{(1+u_1)^2} } \biggl]
    \nonumber\\
    & +  \frac{g_2}{8 \eps},
    \nonumber \\
   Z_3 & =  1 + \frac{g_1\alpha a(1-a) }{(1+u_1)^2 y} +\frac{g_2}{2 \eps},
   \nonumber \\
   Z_4 & =  1 + \frac{g_1}{4(1+u_1)^2 y}\biggl[\alpha
   \NB{1+\frac{4a(1-a)u_1}{1+u_1} } \nonumber\\
   & -  
      u_2(6+6u_1+2\alpha u_1)  \biggl]  + \frac{g_2}{4\eps},
    \nonumber
   \\
   Z_5 & =  1 + \frac{g_1\alpha}{4(1+u_1)^2y}\biggl[1+2(1-a)
   \NB{2a-\frac{1}{1+u_1} }
   \biggl]
   \nonumber\\
   & - \frac{g_1 u_2}{4a(1+u_1)y}\biggl[ 3+\alpha - \frac{2\alpha(1-a)}{1+u_1}
   \biggl] + \frac{g_2(4a-1)}{8a\eps},
   \nonumber \\
   Z_6 & =  1 - \frac{g_1\alpha(1-a) }{(1+u_1) y}\biggl[1-a-\frac{2a}{1+u_1} \biggl]
    + \frac{g_2}{\eps},
    \nonumber \\
   Z_7 & =  1 - \frac{g_1\alpha a}{(1+u_1) y}\biggl[
   a-\frac{2(1-a)}{1+u_1}
   \biggl] + \frac{g_2}{\eps},
   \nonumber \\
   Z_8 & =  1 +\frac{g_1}{2(1+u_1)y}\biggl[\alpha\frac{2a(1-a)+1}{1+u_1} -
   \frac{\alpha a(1-a)}{u_2(1+u_1)^2} \nonumber \\
   & -  u_2(3+\alpha) \biggl] + \frac{g_2}{2\eps}.
\end{align}
The ubiquitous geometric factor stemming from the angular integration
is included into the renormalized charges $g_1$ and $g_2$ via the following
redefinitions
\begin{equation}
   \frac{g_1S_d}{2(2\pi)^d} \rightarrow g_1, \quad
   \frac{g_2S_d}{2(2\pi)^d} \rightarrow g_2,
\end{equation}
where
\begin{equation}
   S_d = \frac{2\pi^{d/2}}{\Gamma(d/2)}
\end{equation}
is the surface area of the unit sphere in the $d-$dimensional space and
$\Gamma(x)$ is Euler's Gamma function.
Equations (\ref{eq:RG_inverse1}) have to satisfy certain conditions
dictated by the symmetry of the model
given by (\ref{eq:time_sym}) and (\ref{eq:supple}). This symmetry results into the
following conditions \cite{AntKap10} for the renormalization constants:
\begin{align}
   \label{eq:RG_cond}
   & Z_i(a) = Z_i(1-a),\quad i\in\{1,2,3,4,8 \} \nonumber \\
   & Z_6(a) = Z_7(1-a), \nonumber\\
   &Z_7(a) = Z_6(1-a), \nonumber \\
   & Z_1(a) - aZ_5(a) = (1-a) Z_5(1-a), 
\end{align}
where the RG constants are considered as functions of the renormalized 
parameter $a$.
By direct inspection of (\ref{eq:RG_inverse1}),
it is easy to see that they indeed fulfill these requirements.

Further,   relations (\ref{eq:RG_direct}) could be inverted with respect
to the RG constants for the fields and parameters in a straightforward manner to yield
\begin{align}
  \label{eq:RG_inverse2}
  Z_D &= Z_2 Z_1^{-1},
  &Z_\tau& = Z_3 Z_2^{-1},\nonumber \\
  Z_v &= Z_4 Z_1^{-1}, 
  &Z_a& = Z_5 Z_4^{-1},\nonumber \\ 
  Z_{\psi} &= Z_1^{1/2} Z_6^{-1/2} Z_7^{1/2},
  & Z_{\tilde{\psi}}& = Z_1^{1/2} Z_6^{1/2} Z_7^{-1/2},\nonumber \\ 
   Z_{u_1} &= Z_1 Z_2^{-1},     
  & Z_{\lambda}& = Z_{1}^{-1/2} Z_2^{-1} Z_6^{1/2} Z_7^{1/2},\nonumber \\ 
   Z_{g_2} &= Z_1^{-1} Z_2^{-2} Z_6 Z_7,
  & Z_{u_2} & = Z_2 Z_8 Z_4^{-2}, \nonumber  \\  
  Z_{g_1} &= Z_2^{-2} Z_4^2. 
\end{align}
After insertion of explicit results for renormalization constants
(\ref{eq:RG_inverse1}), one obtains the desired RG constants of the fields
and parameters of the model.

\section{Anomalous dimensions \label{app:special}}
In this section, we review the explicit expressions
for the anomalous dimension $\gamma_x$, $x\in\{g_1,g_2,u_1,u_2,a\}$ of the
charges and for the fields $x\in\{\psi,\tilde{\psi},\mv\}$, respectively.
From  relations (\ref{eq:RG_inverse2}), the following expressions directly follow:
\begin{align}
    \gamma_D &= -\gamma_1 + \gamma_2, \quad
   & \gamma_a & = -\gamma_4 + \gamma_5, \nonumber\\
    \gamma_{u_2} &= \gamma_2 -2 \gamma_4 + \gamma_8,  \quad
   & \gamma_{\mv} & = -\gamma_1 + \gamma_4,  \nonumber \\
    \gamma_{g_2} &= -\gamma_1 -2\gamma_2 +\gamma_6 + \gamma_7,\quad 
   & \gamma_{g_1} & = 2\gamma_4 - 2\gamma_2,\\
    \gamma_\tau &= \gamma_3 - \gamma_2.
\end{align}
The anomalous dimension $\gamma_x$, corresponding to the renormalization
constant $Z_x, x\in\{1,2,\ldots,8\}$ can be found from the approximate relation
\begin{align}
  \gamma_x & = \mu\partial_\mu \ln Z_x|_0 = (\beta_{g_1}\partial_{g_1}+
  \beta_{g_2}\partial_{g_2}  )\ln Z_x \nonumber \\
   & \approx -(y g_1\partial_{g_1} + \eps g_2 \partial_{g_2})\ln Z_i.
\end{align}
We have subsequently taken into account the following facts:
definitions (\ref{eq:def_gamma}) and (\ref{eq:def_beta}), $Z_i$
could depend only on dimensionless coupling constants 
 and we have retained only the  leading order terms in $\beta$-functions,
which is sufficient in the one-loop approximation. Note that 
$-\eta\D_{u_1}$ has not been included due to the absence of a pole in $\eta$. 
As discussed in literature \cite{Ant99},  this is a property of the low-order 
perturbation theory.
\subsection{General case \label{app:general}}	
The anomalous dimensions for the charges of theory read
\begin{align}
 \gamma_{g_1} & = -\frac{g_1}{2(1+u_1)^2}\biggl[
 3(1+u_1) - u_2(6+6u_1+2\alpha u_1) \nonumber\\
 &+\alpha u_1
 \biggl] - \frac{g_2}{4},\nonumber \\
 \gamma_D & =  \frac{g_1}{4(1+u_1)}\biggl[
      3+\alpha\frac{u_1-1}{u_1+1}+\frac{4\alpha a(1-a)}{(1+u_1)^2}
      \biggl] + \frac{g_2}{8} , \nonumber\\
  \gamma_a & = (1-2a)\biggl[
      \frac{g_1\alpha(1-a)}{2(1+u_1)^3} + \frac{g_1 u_2}{4a(1+u_1)} 
      \biggl( 3+\alpha \nonumber\\
      & -  \frac{2\alpha}{1+u_1} \biggl) + \frac{g_2}{8a}
  \biggl] , \nonumber\\
  \gamma_{u_2} & = 
    \frac{g_1(1-2u_2)}{4(1+u_1)}\biggl[
    3+\alpha\frac{u_1-1}{u_1+1} + \frac{2\alpha a(1-a)}{u_2(1+u_1)^2}
    \biggl]
  -\frac{g_2}{8}, \nonumber\\
  \gamma_{g_2} & = 
    -\frac{3g_1}{2(1+u_1)} + \frac{g_1\alpha}{1+u_1}\biggl[
   \frac{(1-2a)^2}{2} + \frac{1-3a(1-a)}{1+u_1} \nonumber\\
   & + \frac{2a(1-a)u_1}{(1+u_1)^2} 
   \biggl]
   -\frac{3g_2}{2}, \nonumber\\
   \gamma_{\tau} & = -\frac{g_1}{4(1+u_1)}\biggl[
   3+\frac{\alpha}{u_1+1} \biggl(
   u_1-1 + \frac{4a(1-a)}{1+u_1}\biggl)
   \biggl] \nonumber\\
   &- \frac{3g_2}{8}.
    \label{eq:gen_anom_charges} 
\end{align}
In a similar manner anomalous dimensions for the fields can be
computed. The resulting expressions then read
\begin{align}
  \gamma_{\psi} & = \frac{g_1 \alpha}{2(1+u_1)^2}\biggl[
  -a(1-a) + (1+u_1)(2a-1)
  \biggl] - \frac{g_2}{8}, \nonumber\\ 
  \gamma_{\tilde{\psi}} & =
  \frac{g_1 \alpha}{2(1+u_1)^2}\biggl[
  -a(1-a) + (1+u_1)(1-2a)
  \biggl] - \frac{g_2}{8}, \nonumber\\ 
  \gamma_{\mv} & =  \frac{g_1\alpha}{4(1+u_1)^2}\biggl[
     \frac{4a(1-a)}{1+u_1}-1 \biggl] + \frac{g_1 u_2}{2(1+u_1)} \nonumber\\
     & \times 
     \biggl[ 3+ \frac{\alpha u_1}{1+u_1} \biggl]. 
 \label{eq:gen_anom_fields}
\end{align}

\subsection{Rapid-change model\label{app:rapid}}	
Introducing new variables through (\ref{eq:def_rapid}) in relations (\ref{eq:gen_anom_charges}), 
following relations for anomalous dimensions 
\begin{align}
  \gamma_{g_1} & = 
  -\frac{g_1'}{2(1+w)^2}\biggl[
  3(1+w) - u_2(6w + 6 + 2\alpha)+ \alpha\biggl]
  \nonumber \\ 
  &- \frac{g_2}{4}
  ,\nonumber\\
  \gamma_D & = \frac{g_1'}{4(1+w)} \biggl( 3+\alpha\frac{1-w}{1+w} + 
      \frac{4\alpha a(1-a)w^2}{(1+w)^2}   \biggl) +\frac{g_2}{8},\nonumber \\
  \gamma_a & = (1-2a)\biggl(\frac{g_1'\alpha(1-a)w^2}{2(1+w)^3}  + 
      \frac{g_1' u_2}{4a(1+w)}\biggl[  3+\alpha\nonumber\\
      & - \frac{2\alpha w}{1+w}
      \biggl] + \frac{g_2}{8a}   \biggl), 
      \nonumber
      \\
  \gamma_{u_2} & = \frac{g_1'(1-2u_2)}{4(1+w)}\biggl(3+\alpha\frac{1-w}{1+w} +
      \frac{2\alpha a (1-a)w^2}{u_2(1+w)^2} \biggl)
      -\frac{g_2}{8} , \nonumber  \\
  \gamma_{g_2} & = -\frac{3g_1' }{2(1+w)} + \frac{g_1' \alpha}{1+w}\biggl( \frac{(1-2a)^2}{2} +
      w\frac{1-3a(1-a)}{1+w}  \nonumber \\
      & +  \frac{2a(1-a)w}{(1+w)^2}  \biggl)
    - \frac{3g_2}{2},  \nonumber\\
   \gamma_{\tau} & =  \frac{g_1'}{4(1+w)}\biggl[
   3+\frac{\alpha}{1+w} \biggl(
   1-w - \frac{4a(1-a)}{1+w}
   \nonumber\\
   &\times (w+2)w \biggl)
   \biggl] - \frac{5g_2}{8}.
     \label{eq:rapid_anom_charges}  
\end{align}
are obtained. Anomalous dimensions for the fields are given by the expressions
\begin{align}
  \gamma_{\psi} & = \frac{g_1' \alpha}{2(1+w)^2}\biggl[
  -a(1-a)w + (1+w)(2a-1)
  \biggl] - \frac{g_2}{8},\nonumber \\
  \gamma_{\tilde{\psi}} & = \frac{g_1' \alpha}{2(1+w)^2}\biggl[
  -a(1-a)w + (1+w)(1-2a)
  \biggl] - \frac{g_2}{8}, \nonumber \\
  \gamma_{\mv} & =  \frac{g_1' \alpha w}{4(1+w)^2}\biggl(\frac{4a(1-a)w}{1+w} - 1 \biggl) + 
      \frac{g_1' u_2}{2(1+w)} \nonumber \\
      & \times  \biggl(3+ \frac{\alpha }{1+w} \biggl). 
  \label{eq:rapid_anom_fields}
\end{align}

%
\section{Coordinates of the fixed points \label{app:fixed}}
In this section,  we explicitly list analytical expressions for the coordinates of the fixed points.
For  convenience we have introduced a new parameter $a'$ via the relation
$a'=(1-2a)^2$. 
Here NF is an abbreviation for Not Fixed, i.e., for the given FP the corresponding value
of a charge coordinate could not be unambiguously determined. In that case, the given FP
rather corresponds to the whole line of FPs.

The fixed point \fp{II}{6} corresponds actually to the line of possible fixed points
determined by the following system of equations:
\begin{equation}
   g_1^*(1-2u_2^*) = \frac{2y}{3},\quad g_1^*(\alpha {a'}^*-3) = \frac{2y}{3}(\alpha-3).
   \label{eq:append_FP6}
\end{equation}
Further, the coordinates of the last two fixed points 
\fp{II}{7} and \fp{II}{8} are given by the following expressions:
\begin{align}
  g_1^{*} & =  
  \frac{-4}{(\alpha-6) [(\alpha -12) \alpha -180]}
  \biggl[
   (\alpha^2 -12\alpha -72) \eps \nonumber \\
  &+ 3 (21\alpha-2 \alpha^2 +54) y \pm 9 A
  \biggl],
  \nonumber \\
  g_2^{*} & =  -\frac{2 \left[ (21\alpha-2 \alpha^2 +54) y+36 \eps \pm 
  3 A \right]}{(\alpha -12) \alpha-180}, \nonumber \\
  u_2^{*} & =  \frac{4 (\alpha -3) \eps+(42-25 \alpha ) y\pm A}{8 (\alpha -6) 
  \eps -48 (\alpha -3) y},
  \label{eq:nontri_coord}
\end{align}
where $A$ stands for the expression
\begin{align*}
  A& =[-8 (\alpha^2 -9\alpha +126) \eps  y+(49\alpha^2 -372\alpha+1764) y^2
  \nonumber \\
  & + 144 \eps^2]^{1/2}.
  \label{eq:defA}
\end{align*}
The plus sign in (\ref{eq:nontri_coord}) refers to the
point \fp{II}{7}, whereas the minus sign for \fp{II}{8}.

\begin{table*}
  \begin{center}
    \begin{tabular}{|c|c|c|c|c|}
      \hline
      \fp{I}{} & $g_1'^{*}$ & $g_2^{*}$ & $u_2^{*}$ & $a'^{*}$ \\[1.5ex]
      \hline
      \fp{I}{1} & $0$ & $0$ &  NF & NF \\[1.5ex] 
      \hline
      \fp{I}{2} & $0$ & $\frac{2\eps}{3}$&  $0$ & $0$ \\[1.5ex]
      \hline
      \fp{I}{3} & $\frac{4\xi}{3+\alpha}$ &  $0$ & $0$ & NF \\[1.5ex]
      \hline
      \fp{I}{4} & $-\frac{4\xi}{3+\alpha}$ &  $0$ & $\frac{1}{2}$ & $0$\\[1.5ex]
      \hline
      \fp{I}{5} & $\frac{24\xi-2\eps}{3(5+2\alpha)}$ & 
      $ \frac{4\eps(3+\alpha) -24\xi}{3(5+2\alpha)}$&  $0$ & $0$ \\[1.5ex]
      \hline
      \fp{I}{6} & $\frac{2[\eps -4\xi]}{9+2\alpha}$ &
      $ \frac{4\eps(3+\alpha) +24\xi}{3(9+2\alpha)}$  & 
      $\frac{(3+\alpha)\eps -3\xi(7+2\alpha)}{3(3+\alpha)[\eps - 4\xi]}$ & 
      $0 $ \\[1.5ex]
      \hline
      \fp{I}{7} & $-\frac{\xi}{3+\alpha}$ & $2\xi$  & $1$ &
      $-\frac{3(5+2\alpha)}{\alpha}+\frac{2(3+\alpha)\eps}{\alpha\xi}$ \\[1.5ex]
      \hline     
    \end{tabular}
      \caption{List of all fixed points obtained in the rapid-change limit.  The coordinate
  $w^*$ is equal to $0$ for all points.}
       \label{tab:rchm}
  \end{center}
\end{table*}


\begin{table*}
  \begin{center}
    \begin{tabular}{|c|c|c|c|c|c|}
      \hline
      \fp{II}{} & $g_1^{*}$ & $g_2^{*}$ & $u_2^{*}$ & ${a'}^{*}$ \\[1.5ex]
      \hline
      \fp{II}{1} & $0$ & $0$ & NF & NF \\[1.5ex]
      \hline
      \fp{II}{2} & $0$ & $\frac{2\eps}{3}$ & $0$ & $0$ \\[1.5ex]
      \hline
      \fp{II}{3} & $\frac{2y}{9}(3-\alpha)$ & $0$ & $\frac{\alpha}{2(\alpha-3)}$ & 
      $0$ \\[2.5ex]
      \hline
      \fp{II}{4} & $\frac{2 (\eps-y)}{2 \alpha-9}$ & $\frac{4 [3 \eps + 2y ( \alpha- 6)]}
      {2 \alpha-9}$ & 
      $1$ & $\frac{\eps(12-\alpha)+5y(\alpha-6)}{\alpha(\eps-y)}$ \\[1.5ex]
      \hline
      \fp{II}{5} & $-\frac{2 [6 \eps+5 y( \alpha -3)]}{3 (9+\alpha)}$ & $0$ & $
      \frac{3 [\eps+y(\alpha-1)]}{6 \eps+5y  (\alpha -3)}$ & 
           $\frac{18\eps-(\alpha-6)(\alpha-3)y}{\alpha[6\eps+5(\alpha-3)y]}$
      \\[1.5ex]
      \hline      
      \fp{II}{6} & NF & $0$ & NF & NF \\[1.5ex]
      \hline
      \fp{II}{7} & $g_1^{*}$ & $g_2^{*}$ & $u_2^{*}$ & $0$ \\[1.5ex]
      \hline
      \fp{II}{8} & $g_1^{*}$ & $g_2^{*}$ & $u_2^{*}$ & $0$ \\[1.5ex]
      \hline
    \end{tabular}
      \caption{List of all fixed points obtained in the frozen velocity limit. The value
  of the charge $u_1^*$ is equal to  $0$ for all points. }
  \label{tab:fvf}
  \end{center}
\end{table*}

\begin{table*}
  \begin{center}
    \begin{tabular}{|c|c|c|c|c|c|c|}       
      \hline
      \fp{}{} & $g_1^{*}$ & $g_2^{*}$ & $u_1^*$ & $u_2^{*}$ & ${a'}^{*}$ \\[1.5ex]
      \hline
      \fp{II}{7} & $0.532193$ & $9.89135$ & $0$ & $0.37859$ & $0$ \\[1.5ex]   
      \hline
      \fp{III}{1} & $0.365039$ & $6.38225$ & $0.24709$ & $0.352422$ & $0$ \\[1.5ex]      
      \hline
      \fp{III}{2} & $0.399062$ & $7.29847$ & $0.148951$ & $0.35954$ & $0$ \\[1.5ex]      
      \hline
    \end{tabular}     
     \caption{Coordinates of the IR stable fixed points obtained by numerical
  integration of (\ref{eq:invariant_chrg}) for $\alpha=110$ and $\eps = 1$
    in the Kolmogorov regime $y=2\eta=8/3$.
   }
  \label{tab:nontrivial1}
  \end{center}
\end{table*}

\begin{table*}
  \begin{center}
    \begin{tabular}{|c|c|c|c|c|c|c|}       
      \hline
      \fp{}{} & $g_1^{*}$ & $g_2^{*}$ & $u_1^*$ & $u_2^{*}$ & ${a'}^{*}$ \\[1.5ex]
      \hline
      \fp{II}{7} & $0.495405$ & $9.92036$ & $0$ & $0.374461$ & $0$ \\[1.5ex]   
      \hline
      \fp{III}{1} & $0.318124$ & $6.0435$ & $0.32542$ & $0.339525$ & $0$ \\[1.5ex]      
      \hline
      \fp{III}{2} & $0.381096$ & $7.75271$ & $0.121274$ & $0.356122$ & $0$ \\[1.5ex]      
      \hline
    \end{tabular}     
     \caption{Coordinates of the IR stable fixed points obtained by numerical
  integration of (\ref{eq:invariant_chrg}) for $\alpha=110$ and $\eps = 2$
    in the Kolmogorov regime $y=2\eta=8/3$.
   }
  \label{tab:nontrivial2}
  \end{center}
\end{table*}

\begin{table*} 
  \begin{center}
    \begin{tabular}{|c|c|c|c|}       
      \hline
      \fp{}{} & $N(t)$ & $P(t)$ & $R^2(t)$  \\[1.5ex]
      \hline
      \fp{I}{1} & $0$ & $\frac{\eps}{4}-1$ & $1$  \\[1.5ex]   
      \hline
      \fp{I}{2} & $\frac{2\eps}{24-\eps}$ & $\frac{7\eps-24}{24-\eps}$ & $\frac{24}{24-\eps}$  \\[1.5ex]   
      \hline
      \fp{I}{5} & $\frac{(3+\alpha)\eps-6\xi}{3(5+2\alpha)(2-\xi)}$ & 
      $\frac{(18+7\alpha)\eps - 12(5+2\alpha) - 6\xi}{6(5+2\alpha)(2-\xi)} $ & $\frac{2}{2-\xi}$   \\[1.5ex]   
      \hline
      \fp{I}{6} & $\frac{2}{3}\frac{(3+\alpha)\eps-6\xi}{4(9+\alpha)-3(3+\alpha)\eps+2(5+2\alpha)\xi}$ 
      & $\frac{3(\alpha+4)\eps-4(9+2\alpha)-6\xi}{3[4(9+\alpha)-3(3+\alpha)\eps+2(5+2\alpha)\xi]}$  & 
      $\frac{4(9+2\alpha)}{4(9+2\alpha)-3(3+\alpha)\eps+2(5+2\alpha)\xi}$   \\[1.5ex]   
      \hline
      \fp{II}{1} & $0$ & $\frac{\eps}{4}-1$ & $1$  \\[1.5ex]   
      \hline
      \fp{II}{2} & $\frac{2\eps}{24-\eps}$ & $\frac{7\eps-24}{24-\eps}$ & 
		$\frac{24}{24-\eps}$  \\[1.5ex]            
      \hline
    \end{tabular}     
     \caption{Analytical expressions for the given exponents of the Green functions (\ref{eq:scalingGF}). The
     corresponding expressions for FPs \fp{II}7, \fp{III}{1} and \fp{III}{2} are not included, because
     it is not possible to determine their coordinates as explicit functions of free parameters of
     the model, i.e., as $(\eps,y,\eta,\alpha)$. }
  \label{tab:exponents}
  \end{center}
\end{table*}

\bibliographystyle{apsrev}
\bibliography{mybib}

\end{document}